\documentclass[sigconf]{acmart}

\AtBeginDocument{%
  \providecommand\BibTeX{{%
    \normalfont B\kern-0.5em{\scshape i\kern-0.25em b}\kern-0.8em\TeX}}}

\copyrightyear{2025}
\acmYear{2025}
\setcctype{by}
\acmConference[CHI '25]{CHI Conference on Human Factors in Computing Systems}{April 26-May 1, 2025}{Yokohama, Japan}
\acmBooktitle{CHI Conference on Human Factors in Computing Systems (CHI '25), April 26-May 1, 2025, Yokohama, Japan}\acmDOI{10.1145/3706598.3714053}
\acmISBN{979-8-4007-1394-1/25/04}

\usepackage{acmart-taps}
\usepackage{multirow}
\usepackage{listings}
\usepackage{subcaption}
\usepackage{color,soul}
\usepackage{xcolor}
\usepackage{seqsplit}
\usepackage{changepage}
\usepackage{array}

\newcommand{\eg}{\textit{e.g., }}
\newcommand{\cf}{\textit{cf. }}

\newcommand{\ie}{\textit{i.e., }}

\newcommand{\sys}{\textsc{BioSpark}}
\newcommand{\xhdr}[1]{\vspace{.2em}\noindent{{\bf #1.}}}

\newcommand{\revised}[2]{%
  \ifx\relax#1\relax
    \textcolor{black}{#2}%
  \else
    \textcolor{black}{#2}%
  \fi
}

\newcommand{\codett}[1]{{\texttt{\seqsplit{#1}}}}
\newcommand{\code}[1]{{\small\texttt{#1}}}

\definecolor{lightgrey}{HTML}{f0f0f0}
\newcommand{\pquote}[1]{\sethlcolor{lightgrey}\textit{\hl{``#1''}}}
\newcommand{\tind}[3]{$t_{\text{ind.}}$(#1)=#2, $p$#3}
\newcommand{\tpaired}[3]{$t_{\text{paired}}$(#1)=#2, $p$#3}
\newcommand{\chisq}[2]{$\chi^2(1)$=#1, $p$=#2}

\newcommand{\wilcoxon}[2]{$\text{Wilcoxon}$ $W$=#1, $p$=#2}

\newcommand{\pearson}[2]{$\rho$=#1, $p$#2}
\renewenvironment{quote}
  {\list{}{\leftmargin=5pt\rightmargin=5pt}%
    \item\relax}
  {\endlist}

\lstdefinestyle{prompt}{
    basicstyle=\ttfamily\small,
    breaklines=true,
    frame=single,
    breakindent=0pt,columns=fullflexible,
    basewidth=0.95em,
    aboveskip=1em,
    belowskip=1em,
    emphstyle=\textit,
}

\lstdefinestyle{code}{
    basicstyle=\ttfamily\small,
    breaklines=true,
    frame=single,
    breakindent=0pt,columns=fullflexible,
    basewidth=0.95em,
    aboveskip=1em,
    belowskip=1em,
    emphstyle=\textit,
}
\newcommand\cirnum[1]{\raisebox{.5pt}{\textcircled{\raisebox{-.9pt}{#1}}}}

\newcommand{\rev}[1]{\textcolor{black}{#1}}

\begin{document}

\title[\sys]{\sys: Beyond Analogical Inspiration to LLM-augmented Transfer}

\author{Hyeonsu B. Kang}
\email{hyeonsuk@cs.cmu.edu}
\orcid{0000-0002-1990-2050}
\affiliation{%
  \institution{Human-Computer Interaction Institute}
  \streetaddress{Carnegie Mellon University}
  \city{Pittsburgh}
  \state{PA}
  \country{USA}
  \postcode{15213}
}

\author{David Chuan-en Lin}
\affiliation{%
  \institution{Human-Computer Interaction Institute}
  \streetaddress{Carnegie Mellon University}
  \city{Pittsburgh}
  \state{PA}
  \country{USA}
  \postcode{15213}
}
\email{chuanenl@cs.cmu.edu}

\author{Yan-Ying Chen}
\affiliation{%
  \institution{Toyota Research Institute}
  \city{Los Altos}
  \state{CA}
  \country{USA}
  \postcode{94022}
}
\email{yan-ying.chen@tri.global}

\author{Matthew K. Hong}
\affiliation{%
  \institution{Toyota Research Institute}
  \city{Los Altos}
  \state{CA}
  \country{USA}
  \postcode{94022}
}
\email{matt.hong@tri.global}

\author{Nikolas Martelaro}
\affiliation{%
  \institution{Human-Computer Interaction Institute}
  \streetaddress{Carnegie Mellon University}
  \city{Pittsburgh}
  \state{PA}
  \country{USA}
  \postcode{15213}
}
\email{nikmart@cs.cmu.edu}

\author{Aniket Kittur}
\affiliation{%
  \institution{Human-Computer Interaction Institute}
  \streetaddress{Carnegie Mellon University}
  \city{Pittsburgh}
  \state{PA}
  \country{USA}
  \postcode{15213}
}
\email{nkittur@cs.cmu.edu}

\renewcommand{\shortauthors}{Kang et al.}

\begin{abstract}
We present \sys{}, a system for analogical innovation designed to act as a creativity partner in reducing the cognitive effort in finding, mapping, and creatively adapting diverse inspirations.
While prior approaches have focused on initial stages of \rev{finding} inspirations, \sys{} uses LLMs embedded in a familiar, visual, Pinterest-like interface to \rev{go beyond inspiration to} supporting users in \rev{identifying the key solution mechanisms, transferring them to the problem domain, considering tradeoffs, and elaborating on details and characteristics. To accomplish this \sys{} introduces several novel contributions}, including a tree-of-life enabled approach for generating relevant and diverse inspirations\rev{, as well as AI-powered cards including} `Sparks' for analogical transfer; `Trade-offs' for \rev{considering pros and cons}; and `Q\&A' for deeper \rev{elaboration}. We evaluated \sys{} through workshops with professional designers and a controlled user study, \rev{finding that using \sys{} led to a greater number of generated ideas; those ideas being rated higher in creative quality; and more diversity in terms of biological inspirations used than a control condition. Our results suggest new avenues for creativity support tools embedding AI in familiar interaction paradigms for designer workflows.}
\end{abstract}

\begin{CCSXML}
<ccs2012>
   <concept>
       <concept_id>10003120.10003121.10003129</concept_id>
       <concept_desc>Human-centered computing~Interactive systems and tools</concept_desc>
       <concept_significance>500</concept_significance>
       </concept>
   <concept>
       <concept_id>10003120.10003121.10003128</concept_id>
       <concept_desc>Human-centered computing~Interaction techniques</concept_desc>
       <concept_significance>500</concept_significance>
       </concept>
   <concept>
       <concept_id>10003120.10003121.10003122</concept_id>
       <concept_desc>Human-centered computing~HCI design and evaluation methods</concept_desc>
       <concept_significance>500</concept_significance>
       </concept>
   <concept>
       <concept_id>10002951.10003260</concept_id>
       <concept_desc>Information systems~World Wide Web</concept_desc>
       <concept_significance>500</concept_significance>
       </concept>
   <concept>
       <concept_id>10002951.10003317.10003331</concept_id>
       <concept_desc>Information systems~Users and interactive retrieval</concept_desc>
       <concept_significance>500</concept_significance>
       </concept>
 </ccs2012>
\end{CCSXML}

\ccsdesc[500]{Human-centered computing~Interactive systems and tools}
\ccsdesc[500]{Human-centered computing~Interaction techniques}
\ccsdesc[500]{Human-centered computing~HCI design and evaluation methods}
\ccsdesc[500]{Information systems~World Wide Web}
\ccsdesc[500]{Information systems~Users and interactive retrieval}

\keywords{Goal-driven Analogies, Analogical Transfer, Design Creativity, Ideation, Large Language Models}

\begin{teaserfigure}
\centering
    \includegraphics[width=0.8\linewidth]{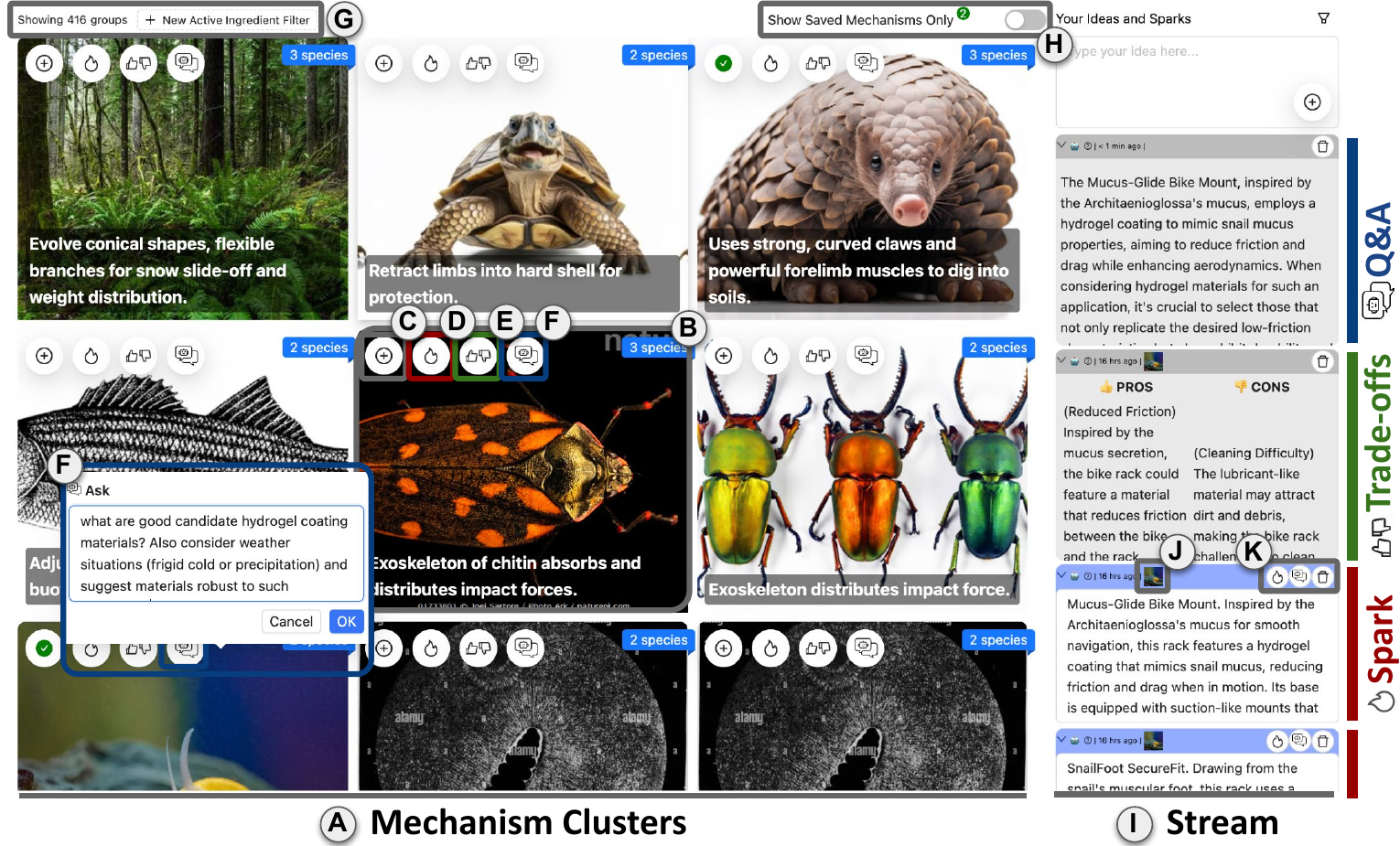}
    \vspace{-6pt}
    \caption{The \sys{} interface consists of \textit{mechanism active ingredient} clusters (\cirnum{A}) and a stream panel (\cirnum{I}). Cluster cards (\cirnum{B}) display species images, descriptions, and action buttons, enabling users to save mechanisms (\cirnum{C}), generate \textit{Sparks} (\cirnum{D}), explore \textit{Trade-offs} (\cirnum{E}), and use a \textit{Q\&A} chat (\cirnum{F}). The stream panel (\cirnum{I}) contains system- and user-generated ideas, which are interactive and editable (Details in \S\ref{section:biospark}).}
    \label{fig:interface}
\vspace{-6pt}
\end{teaserfigure}

\maketitle

\section{Introduction}
Many innovations in design, technology, and science have been driven by people finding and adapting analogical inspirations from fields distant to their own.
\rev{\textit{Analogical innovation} refers to the process of applying insights from one domain to solve problems or create new designs in another.}
Whether Vetruvius explaining how sound waves work through analogy with water waves~\cite{darrigol2010analogy}, the Wright brothers designing a lightweight wing control mechanism based on a bicycle inner tube box~\cite{johnson2005flying}, or engineers partnering with an origami expert to furl a solar array into a narrow rocket~\cite{Monk2013OrigamiSpace,peraza2014origami,zirbel2013origami}, such innovations have required their inventors to deeply engage in a complex cognitive process of finding and creatively adapting inspirations that had limited surface similarities but deep structural similarities.

While such analogies may sometimes seem like `lightning strikes' of serendipity, researchers have identified that analogical innovation involves several cognitive stages \rev{of processing, each} of which can require significant mental effort ~\cite{gickholyoak1980,gentner1983structure,gentner1985analogical}.
First, \rev{\textbf{finding} }potential inspirations in distant domains is difficult because of the challenge of \rev{identifying} domain\rev{s and inspirations that might contain useful potential analogical mechanisms.
However, even when potential inspirations are found, significant cognitive effort is needed as part of the broader process of \textit{analogical processing} to \textbf{recognize} and \textbf{understand} what the \textit{active ingredients} of the inspiration's mechanisms are, such as key structural or functional properties that enable its utility (\eg the shearing properties of the cardboard inner tube box).
Once identified, another barrier involves exploring how to map and \textbf{transfer} these mechanisms to the target problem (a process known as \textit{analogical transfer}).
For example, instead of four sides of a box, using a set of cables to create shearing in two parallel planes of the wings).
Finally, the mechanisms  need to be \textbf{adapted} to consider tradeoffs and limitations in the target domain  (\eg whether cables for connecting the wings could provide a sufficient balance of rigidity and weight reduction for steering the wings through shearing).}

Supporting these complex needs in a single system has been challenging, with most \rev{past} approaches focusing on one or two stages of the process and largely limited to a small, hand-coded set of inspirations~\cite{jiang2022data,asknature,DANE,SAPPhIRE}.
Approaches to collecting inspirations at a larger scale have begun to appear~\cite{analogy_search_engine,hope_kdd17,huang2020coda,emuna2024imitation}, but have mostly been limited to helping with the finding stage of analogical innovation.
Relying on users to do the hard work of determining which inspirations could be relevant and how they could be adapted can lead to them not noticing or putting in the effort to go beyond surface similarities and try to understand how an inspiration could be used; as noted in Kang et al. 2022, ``the critical first step towards analogical inspiration may be raising... enough attention and interest above the initial `hump' of cognitive demand''~\cite{analogy_search_engine}.

In \sys{}, we explore the idea of a LLM-powered computing system acting as a creativity partner to proactively help with the intellectual work of not only finding analogies but also \rev{mapping,} transferring, and adapting those ideas to the target domain.
By doing so we aim to help free up the cognitive effort of users to engage in the creative process of exploring new design spaces and considering more ideas more deeply than they would be able to otherwise.

To achieve this, \sys{} explores several new design patterns for partnering AI with human analogical ideation, including:
\begin{itemize}
    \item A tree-of-life enabled approach for generating new and relevant biological inspirations from a small set of `gold standard' inspirations taken from AskNature;
    \item An analogical ideation interface leveraging familiar interaction concepts from designers' practice of browsing Pinterest and curating moodboards \rev{that helps them recognize the active ingredients of the inspirations' mechanisms};
    \item Proactively generating `sparks' that help users understand the mapping \rev{and transfer} between inspirations and their design problem;
    \item Providing pro/con trade-offs to scaffold users in considering \rev{how to adapt} key aspects of the design problem;
    \item Supporting a free-form Q\&A interface grounded in the inspiration and the design problem context to help users more deeply consider \rev{and elaborate on} inspiration mechanisms.
\end{itemize}

We instantiated \sys{} in a prototype system and evaluated and iterated on it through a workshop study with 6 professional designers, a formative study with 4 participants with design and engineering backgrounds as well as a user study with 12 participants of varied backgrounds. Our results suggest \rev{novel} ways in which AI support can be embedded into interfaces to support and augment human creativity.

\section{Related Work}
\subsection{Design by analogy}
Throughout history, analogies have often driven breakthroughs in science, engineering, and design (\eg~\cite{darrigol2010analogy,johnson2005flying,Monk2013OrigamiSpace}).
Yet, analogical innovation in human minds has proven rare due to the cognitive challenges involved with the underlying analogical processing.
One challenge is the high sensitivity to surface-level similarity during retrieval from memory that favors analogs with shared visual or keyword similarities over the ones that share a deeper underlying structure~\cite{gentner1985analogical}.
In addition, the heavy cognitive load incurred during analogical processing, even with just a few relations, significantly burdens working memory and leads to performance degradation~\cite{gick1983schema,gentner1993roles,halford2005many}. 
To support people with analogical processing, researchers have designed various systems for analogy retrieval.
One thread of research here focuses on modeling analogical relations, albeit in limited scopes.
This includes system based on the structure-mapping theory~\cite{gentner1983structure,forbus1994incremental,forbus2001exploring}, multiconstraints theories (\eg~\cite{holyoak1989analogical}, connectionist designs~\cite{hummel2003symbolic,hofstadter1995copycat}, and rule-based approaches~\cite{ashley1991reasoning,carbonell1985derivational,carbonell1983learning}).
Many methods involve labor-intensive processes, such as the WordTree methodology~\cite{linsey2012design}.
Additionally, numerous systems depend on hand-coded and meticulously structured data, the curation of which is often resource-intensive (\eg~\cite{DANE,BioTRIZ}).

Recent work in computational methods for finding analogical inspirations at scale have shown promising results using a significantly simplified schema (\eg the purpose and mechanism schema in~\cite{analogy_search_engine,hope_kdd17}) with just a fraction of data (\eg~\cite{analogy_search_engine,hope_kdd17,chan2018solvent}).
However these systems primarily focus on facilitating the discovery of potential analogies and do not extend support to the subsequent, intricate stages of design that follow.
This involves navigating potential limitations or trade-offs, which are essential for the successful transfer of these analogies in real-world scenarios~\cite{ashby2013materials,idea_inspire,BioTRIZ}.

\subsection{Bioinspired design}

One particularly relevant thread of research in design-by-analogy focuses on finding inspirations in biological organisms and systems \cite{jiang2022data}. However, prior approaches have been limited due to their reliance on costly manual curation (\eg AskNature~\cite{asknature} or DANE~\cite{DANE}; for example, the researchers of DANE found that re-describing a single biological organism in the Structure-Behavior-Function framework can take approximately $\sim$40-100 hours per model).
Alternative approaches demonstrated the feasibility of using crowdsourcing to power supervised learning for identifying scientific articles with biomemetic inspirations (\eg~\cite{IBM,biologue}), but the cost of curating high-quality annotations presented a significant bottleneck for scalability.
Yet another line of research has explored rule-based (\eg~\cite{Cheong2014}) or data programming~\cite{emuna2024imitation} approaches, and showed promising results, albeit potential concerns of their generalizability and scalability.

Our iterative tree-of-life-based algorithm for expanding the mechanism dataset builds on these threads of research, while also leveraging recent advances in AI, such as Large Language Models (LLMs), that present promising new opportunities for designing scalable approaches for bio-analogy generation.
However, naively prompting LLMs in a zero-shot manner may still result in limited diversity on abstract concepts~\cite{chung_2023_llm_diversity}.
One possible avenue of research here is to further exploring knowledge-augmented or knowledge-guided prompting, which has been previously explored in factual Q\&A for improving the factuality in answers to simple questions  (\eg ``\textit{Where did Alex Chilton die?}'') by traversing a knowledge graph known to contain relevant facts to contextualize LLM prompts, for the goal of increasing the conceptual diversity in generation.
Another related thread here is recent work on self-feedback and refinement techniques~\cite{madaan2024self}, that showed iteratively generating feedback on the LLM output using the same LLM and refining based on it leads to better performance.
Our tree-of-life enabled approach takes inspirations from these prior works to develop an algorithm that uses the knowledge structure for diversification.

\subsection{Creativity Support Systems (CSTs)}
\sys{}, in its goal of developing a creativity partner for analogical innovation, also builds on the related work in Creativity Support Systems (CSTs). Specifically, co-creativity~\cite{mamykina2002collaborative} and human-AI collaborative systems have been explored in the CST literature, leading to related systems such as Drawing Apprentice~\cite{davis2015drawing}, Due-Draw~\cite{oh2018lead}, Creativity Sketching Partner~\cite{karimi2020creative}, and Collaborative Ideation Partner~\cite{kim2023effect} which aimed to produce useful output for building on human user input with a joint similarity to both the design task and the input. The goal-driven approach in \sys{} also focuses on the relevance to the task context, but differs from these works in its emphasis on analogical relevance which relies on structural rather than feature similarity.

Another related thread of research in the CST literature is systems that support analogies as an ideation method, as seen in related works such as MetaMap~\cite{kang2021metamap}, WikiLink~\cite{zuo2022wikilink}, Idea-Inspire~\cite{idea_inspire}. \rev{These works focus on the initial stages of helping users find relevant analogies, but stop short of helping users in the cognitive process of analogical transfer.} While transfer has been studied as a critical failure point in analogy for decades in cognitive psychology~\cite{gick1983schema}, \rev{it remains understudied in} the CST literature, \rev{perhaps} due to the difficulty of even finding analogical inspirations, let alone \rev{addressing how to transfer} them to the target domain.
\sys{} contributes to this gap in the literature by helping designers recognize, transfer, explore and elaborate on new design spaces through analogies and their example instantiations, which~\cite{lee2024and} identified as ``grounded metaphors'' and highlighted as open research areas.

\subsection{LLMs for ideation and co-creation}

Recent advances in LLMs also suggest the potential for scalably augmenting analogical innovation for users throughout the entire cognitive process, from finding potential analogical inspirations to mapping them to the problem domain to helping users more deeply engage with their mechanisms and trade-offs.
LLMs have shown the capability to infer specific analogies and to generate ideas relevant to a design goal (\cf~\cite{webb2023emergent}).
They also can serve as more flexible natural language processing components in an interface, allowing for powerful interface augmentation approaches (\eg~\cite{KANG2023Synergi,liu2023selenite,fok2023qlarify,august2023paper} as well as direct interfaces using chat-based dialog (\eg~\cite{MicrosoftCopilot2023,AskYourPDF,OpenAI_ChatGPT}).

However, studies examining the use of LLMs and generative AI in the creative process have shown that improperly incorporating LLMs into the creative process can end up doing more harm than good.
Using generative AI systems such as image generation (\eg Midjourney) or text generation (\eg ChatGPT) has been shown to lead users to become more fixated rather than more creative \cite{wadinambiarachchi2024effects}.
Several core properties to LLMs have been identified as potentially problematic, including tendencies for inaccurate inferences and hallucinations, user fixation on the initial prompts they enter, and overly accepting the results of AI-generated ideas rather than adapting them or using them to further explore the design space~\cite{wadinambiarachchi2024effects,kim2023effect}.
These results suggest a more nuanced approach to incorporating LLMs and AI into the analogical innovation process may be needed.

\section{Formative Studies and Design Goals}
We developed \sys{} in an iterative process involving an initial formative workshop study with professional automobile designers for investigating their current workflows of getting inspirations for concept design, followed by a design probe with additional designers using a partial development of the initial \sys{}'s goal-driven data discovery pipeline which generated a set of analogical mechanism inspirations (\S\ref{subsection:data_discovery_diversification_structure}).

\rev{Our initial day-long workshop included six professional designers from a large multinational company specializing in mobility design. The workshop provided insights into how our system could integrate into their daily workflows, highlighting basic design requirements such as the \textbf{importance of visual, goal-centered inspirations} and \textbf{leveraging familiar interfaces} for a system that could be incorporated into their practice, which provided constraints for the development of \sys{}.}

Informed by the workshop findings, we developed a design probe to further investigate challenges in AI-augmented bioinspired design. The probe consisted of a set of images of inspirations arranged in a moodboard or Pinterest-like visual style that users could scroll through.  Inspirations were sampled from a prototype of the inspiration generator we describe later, and focused on two design briefs: designing a secure bike rack and improving driving on slippery roads.  This included mechanisms ranging from geckos climbing walls to slime lubrication.  Each individual mechanism had an associated visual image retrieved using online search (see Appendix~\ref{appendix:design_inspiration_probe_image_search}). To help designers experience what potential AI assistance could feel like we included prototype functionality for designers to interact with these inspirations (\ie buttons that would prompt an LLM to provide an additional explanation, or to compare or combine a pair of mechanisms; see more in Appendix~\ref{appendix:design_inspiration_probe_interface}).  

We recruited four designers (none of whom participated in the earlier workshop) with backgrounds in design and engineering to provide feedback as pilot participants. Participants found aspects of the prototype valuable, and all did find mechanisms that inspired new design ideas (e.g., the coiling of octopus tentacles and lizard tails inspiring bike rack components that could expand and contract with turbulence; or scale and fur arrangements in rodents inspiring groove patterns on tires that would create more downforce on slippery roads). 

However, they also brought up several fundamental challenges with the approach that motivated \sys{}.  In particular, all participants noted a similar theme in terms of the challenges involved with \textbf{recognizing and understanding }the `active ingredients' (\ie the core abstraction underpinning how each mechanism actually works) of inspirations with respect to the design problem. 
For example, it could also be difficult to understand how the textual descriptions of individual mechanisms were relevant to the design problem (\eg \revised{}{\pquote{I want to choose an interesting animal that can support force and dynamics... but there are a lot of bugs, birds, and dolphins, not sure how to think about them as relating to turbulence reduction} -- P3).}

Participants also wanted additional support for envisioning how inspirations would \textbf{transfer into target design domains}.
Participants commented that assisting them with applying active ingredients to focused areas in the target domain could help with generating new design ideas based on distant analogies \revised{}{\:\pquote{`slime secretion' as relating to the attachment/detachment mechanism and the friction aspect rather than the slime itself would be more helpful'} -- P1; \pquote{Maybe it (the system) can tell me about skin texture of frogs for generating ways to modulate it and manage turbulence?} -- P2)}.

Finally, participants noted the desire for \textbf{deeper elaboration on inspirations}. This included understanding benefit and drawbacks of the mechanism with respect to dimensions relevant to its effectiveness in the problem domain, as in this quote from P4: \pquote{A few words to highlight the most interesting properties or dimensions for efficiency, durability, or versatility would be very helpful. I want to get a bigger picture here} (P4). It also included understanding more details about a particular mechanism, as commented by P1: \pquote{`Bike rack' and `slime' are somewhat contradictory but it (slime mechanism) makes me think about the attachment aspects of the design... maybe new ideas around loading and unloading of bikes that have dynamically adjusting surface friction... I'm going to click ``explain'' on slime... (after the detailed explanation loads) I wish I could know more about the lubrication mechanism aspect of slimes}. 
\vspace{-1em}

\section{\sys{}}
\label{section:biospark}

We instantiated these design goals in \sys, with the high level intention of acting as a creative partner in the analogical design and innovation process beyond simply finding inspirations to proactively helping the user \rev{understand the key solution mechanisms in those} inspirations\rev{, how to transfer them to} their own problem domain, \rev{and to more easily} consider their trade-offs \rev{on the problem's} design constraints or \rev{elaborate on} more information about their details and characteristics.

\rev{To support these functions in a simple and familiar interaction} the AI provides its suggestions on mappings between inspirations and the problem as idea `spark' cards that are added to a sidebar whenever the user saves an inspiration; `trade-off' cards that contextualize the pros and cons of an inspiration's mechanism within the problem domain; and Q\&A cards that allow the user to submit free-form queries to the LLM which are automatically contextualized with the problem and inspiration contexts.
As the user engages with the system a typical flow involves them perusing and saving inspirations, and engaging with the cards in the sidebar to more deeply consider particular mechanisms or the design spaces they represent.

In the following sections we describe in more detail the design of the system, starting with a \revised{}{a detailed description and a scenario walkthrough of the system.}

\subsection{Scenario}
 \label{subsubsection:walk_through}
Consider an automotive designer, Sarah, looking for inspirations that could spark new ideas for novel bike rack design.
When she arrives at the \sys{} interface, she first scrolls through the board UI on the left of the screen to review different clusters of mechanism \revised{}{active ingredients}.
She is initially \revised{drawn to}{intrigued by} the `exoskeleton' cluster, \revised{showing}{which shows} an image of a froghopper, \revised{as}{because} the exoskeleton structure may \revised{have}{provide} insights \revised{into the skeletal design of new bike racks}{for new skeletal bike rack designs}.
She clicks the cluster card (fig.~\ref{fig:interface}, \cirnum{A}) to examine its details further.
The mechanism description in the modal that expands out \revised{upon}{on} her click highlights `chitin', as strong \revised{and}{yet} flexible material, that can absorb and distribute the force of impact.

She then finds another mechanism that seems counter-intuitive yet interesting, the mucus and muscular foot of `Architaenioglossa' \revised{}{-- an order of snails --} as potentially interesting \revised{mechanisms}{active ingredients} for the problem.
She clicks on the `spark' button to receive inspirations for \revised{new ideas that may use this mechanism in new ways}{creatively adapting the active ingredients} (fig.~\ref{fig:interface}, \cirnum{D}).
She receives two sparks in response; the first, titled `Mucus-Glide Bike Mount', describes an idea that uses hydrogel coating to reduce friction in motion.
Intrigued by the idea, but concerned with the durability of hydrogel in various weather conditions, she asks \sys{} using the `Q\&A' button (fig.~\ref{fig:interface}, \cirnum{F}): ``\textit{what are good candidate hydrogel coating materials? Also consider weather situations (frigid cold or precipitation) and suggest materials robust to such conditions.}''.
\sys{} returns an information card that provides alternative material choices, such as Polyacrylamide Hydrogels, described as capable of maintaining their mechanical strength and elasticity in a wide range of temperatures and as resistant to degradation in wet conditions, or Polyvinyle Alcohol (PVA) Hydrogels, notable for excellent mechanical properties and withstanding repeated freeze-thaw cycles while maintaining a low-friction surface even when wet, which makes them an appealing case for use in cold weather conditions (the Q\&A card in the top of the stream, fig.~\ref{fig:interface}, \cirnum{I}).
She writes down these materials as potential leads to pass on to the engineering research team later, and clicks on the `Trade-off' button (fig.~\ref{fig:interface}, \cirnum{E}) to learn more about the potential disadvantages of a design that incorporates a lubricant-like material directly on the surface of the rack where bike wheels are loaded on to.
The returned trade-offs card raises cleaning difficulty as a potential concern, which she uses to ideate related usage scenarios and constraints involved to develop the idea further.

\begin{figure}[h]
    \centering
    % \vspace{-1em}
    \includegraphics[width=\linewidth]{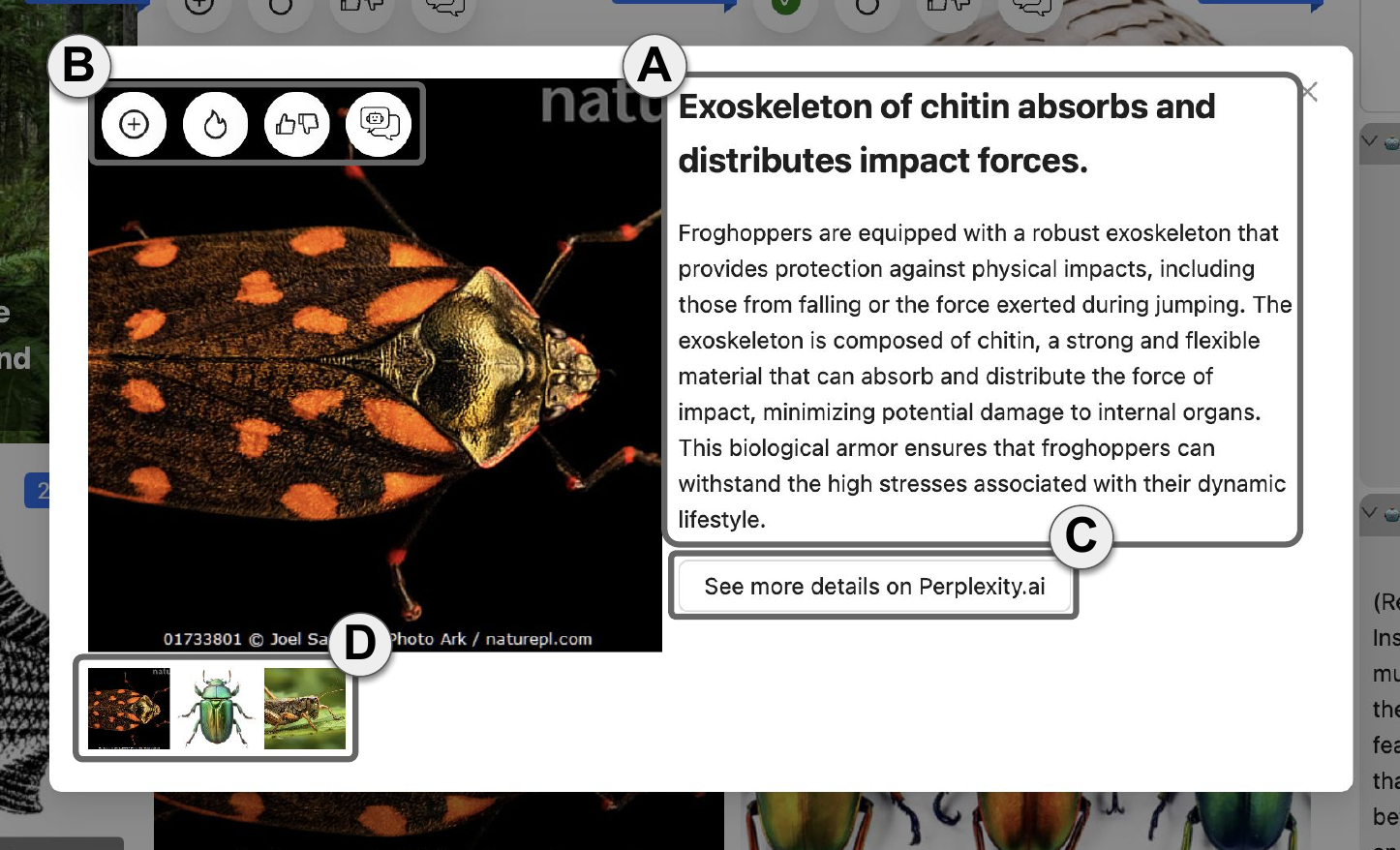}
    \vspace{-1em}
    \caption{The modal view of a clicked mechanism cluster shows additional mechanism and active ingredient details (\cirnum{A}). The same action buttons featured on the main page of the interface (\cirnum{B}) are shown, as well as the `See more details on Perplexity.ai' for finding additional details and related scientific researech (\cirnum{C}), and a carousel displaying other species that belong to the cluster which can be viewed by clicking on any of the images (\cirnum{D}).}
    \vspace{-2em}
    \label{fig:modal}
\end{figure}

\subsection{Leverage familiar interfaces and behavior}
Consistent with findings from the design workshop we aimed to design a system that could leverage familiar interface paradigms and existing behaviors in new ways. We based our overall design on the concept of a mood board, similar to Pinterest, where designers normally scroll through and collect many possible inspirations.
To this end we added a timeline-style side pane that could act as a central focus for AI suggestions and as a scratch and organization space for the user to externalize and keep track of their thinking and ideation process separate but related to their foraging for inspirations.

To support the side pane's use as an organization or triage space we also enable quick filtering of different types of AI-provided information cards (see fig.~\ref{fig:interface}, top of the stream \cirnum{I}), as well as deleted items with support for restoration.
Users can build off any cards in the stream to explore the design space in a non-linear fashion, with that spark anchoring context for addition exploration.

\sys{} was implemented using \codett{React.js} for the interface and the \codett{Flask} server in \codett{Python3.11} for the backend components.

\subsection{Visual representation}
\label{subsection:mechanism_image_retrieval}
To support the design goal of visual representations of inspirations we were inspired by AskNature.org, a popular web repository for bio-inspired design.
AskNature pages are designed to include a prominent close-up and centered portrait of a species that creates a striking visual and invokes curiosity. 

To achieve this we used Google Search and Adobe Stock Images for retrieving relevant inspiration images. Directly searching on Google using its API with an animal name or mechanism description query often resulted in images such as book covers or graphs in related research paper, which were less effective.
Conversely, using Adobe Stock Images\footnote{\url{https://stock.adobe.com}} with animal species names as queries sometimes led to high quality photos but other times the animal was shown in the distance or background, and coverage was limited. Therefore, for each species we combined the top-5 results from Google Search and Adobe Stock Images to create a set of images, and used GPT-4V (\codett{gpt-4-vision-preview}) to rank each of the species' images in terms of the visual focus and the potential value for mechanism understanding (fig.~\ref{fig:ranking_species_images_ex}; full prompt in Appendix~\ref{appendix:ranking_species_images} and the score and rationale of each image in Appendix~\ref{appendix:sample_scores_and_justifications}).
The highest-scoring image was chosen to represent each species, and the results appeared sufficiently accurate for the needs of the prototype system.

\subsection{Goal-driven inspiration discovery}
\label{subsection:data_discovery_diversification_structure}
\begin{figure*}[h!]
    \centering
    \vspace{-1em}
    \includegraphics[width=\linewidth]{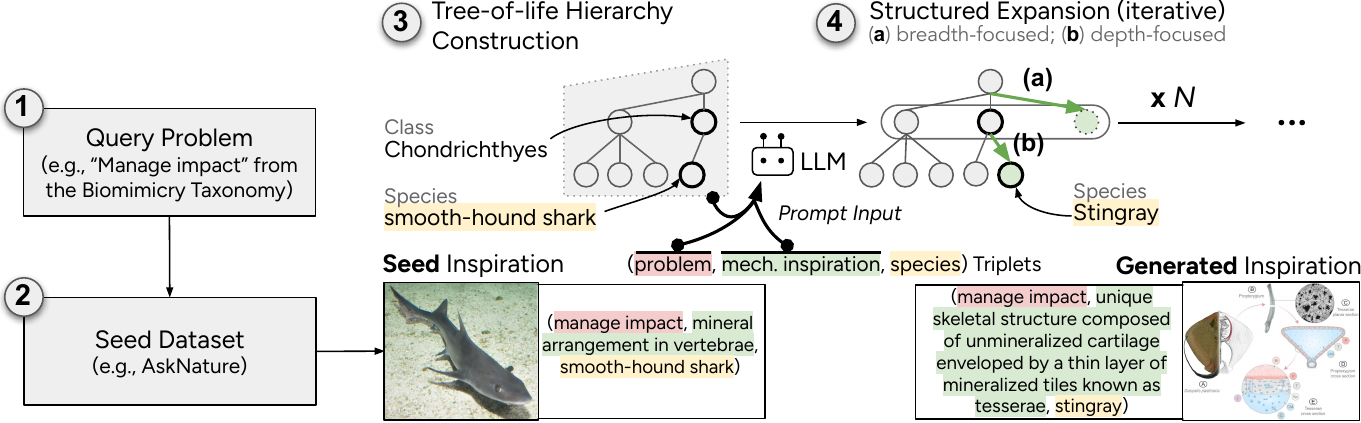}
    \vspace{-2em}
    \caption{\rev{\textbf{Goal-driven Mechanism Inspiration Generation Pipeline}. The pipeline begins from \cirnum{1} a query problem, such as a function `Manage Impact' from the BioMimicry Institute's Taxonomy which covers a broad range of problems. The problem is used to \cirnum{2} search AskNature.org, which organizes species or innovations according to the functions, to provide seed data. \cirnum{3} This data is structured into the (problem, mechanism, species) schema and the species names are used to prompt an LLM to construct the tree-of-life hierarchy. \cirnum{4} The hierarchy is traversed to identify expansion sites at the frontier, here instantiated as sparse branches on the tree with high diversification opportunities, determined in the breadth- or depth-focused manner. The sites along with the problem context and existing mechanisms in the dataset are provided to contextualize the generation and to jointly enforce diversification and relevance filtering. The generation continues iteratively until a stopping condition is met. While the pipeline uses specific data sources in this example, the approach is source-agnostic and adaptable to other contexts (see text).}}
    \vspace{-1em}
    \label{fig:backend_tree_generation}
\end{figure*}
Finding biological inspirations is a longstanding research challenge, as discussed in the related work.
This challenge is made more difficult when driven by a specific design goal, which might be relevant to only a tiny proportion of possible biological inspirations.
Instead of searching for needles in a haystack (\eg through various means such as crowdsourcing~\cite{IBM,biologue}, rule-based programs such as~\cite{Cheong2014}, data programming~\cite{emuna2024imitation}) we explore the idea of leveraging recent advances in language model training to use LLMs as a knowledge retriever.
However, in comparison to existing approaches such as fine-tuning with a set of biological examples (\eg from AskNature~\cite{zhu2023biologically}) with no resulting control over the areas of the design space explored or the diversity of the resulting inspirations found, we introduce an approach that introduces both structure and diversification.

Our approach conceptually follows the hierarchy-based expansion mechanism from the WordTree method~\cite{linsey2012design} that demonstrated how up-then-down traversal on an abstraction hierarchy in structured brainstorming settings could lead to novel insights.
Here, we design a similar approach for structurally expanding a seed dataset (in our case, AskNature biological inspirations) but instead of an arbitrary abstraction hierarchy we use a structured hierarchy latent to most biological inspirations: the Tree-of-Life\footnote{\url{https://en.wikipedia.org/wiki/Tree_of_life}} model.
This approach anchors our expansion algorithm on species of nature as a mediator for exploring new spaces of mechanisms, as species often adapt to changing natural environments by evolving with new mechanisms. 

\rev{Concretely, starting from a query problem `manage impact', an example seed mechanism found on AskNature.org may be the mineral arrangement in smooth-hound shark vertebrae post}\footnote{\url{https://asknature.org/strategy/minerals-strengthen-vertebrae/}} (fig.~\ref{fig:backend_tree_generation}).
\rev{We structure this into a triplet of (problem, mechanism inspiration, species): (`manage impact', `mineral arrangement in smooth-hound shark vertebrae', smooth-hound shark) (details in Appendix}~\ref{appendix:schemas_from_asknature}).
\rev{From the seed triplets, we first construct a 7-level tree-of-life hierarchy using GPT4 consisting of the \codett{\{domain, kingdom, phylum, class, order, family, genus, species}\} levels by prompting it with the name of the species.
In our evaluation using 90 `ground-truth' taxonomies (sourced from Wikipedia's `biota' scientific classification information boxes), we find satisfactory accuracy levels ranging between 94.4\% -- 100\% for each level on the taxonomy} (Table~\ref{table:taxonomy_generation_accuracy}, Appendix~\ref{appendix:accuracy_taxonomy_construction}).

\rev{We then traverse the constructed tree-of-life hierarchy to identify the directions of expansion that have high potential for diversification.
We design two approaches -- breadth- and depth-focused expansions -- to this end, and estimate the potential for diversification as sparsely populated branches on the hierarchy, as expanding from them is likely to result in new species that are not currently represented in the dataset.}

\rev{In breadth-focused expansion, we design a prompt} (fig.~\ref{fig:breadth_expansion_prompt} in Appendix~\ref{appendix:iterative_structured_expansion}) \rev{that requests GPT4 to generate sibling nodes of a given node, excluding the previously generated nodes to avoid duplicate generation.
In depth-focused diversification, we filter nodes at a given level (\eg `order') and select the five with the fewest children (those having the greatest opportunity for vertical exploration), and design a prompt} (fig.~\ref{fig:depth_expansion_prompt} in Appendix~\ref{appendix:iterative_structured_expansion}) \rev{to prompt GPT4 for children nodes generation.}

To maintain the relevance to the query problem, we add the descriptions of the query problem along with specific constraints to the prompt to contextualize the generation and to jointly enforce diversification and relevance filtering.
After this process the system generates new mechanism inspirations.
Using the seed triplet (`manage impact', `mineral arrangement in smooth-hound shark vertebrae', smooth-hound shark) as an example, one example of the newly generated mechanism inspirations may be the unique skeletal structure composed of unmineralized cartilage enveloped by a thin layer of mineralized tiles known as tesserae in stingrays.
Both stingrays and gray smooth-hounds are species of the class of jawed fish Chondrichthyes.
The newly generated data is incorporated back to the tree-of-life hierarchy updating its frontier, and the algorithm runs iteratively until a stopping condition is met.

\xhdr{Approach Generalizability}
\rev{While the pipeline uses specific data sources here, the general approach is agnostic to the source of data and can be applied to any sources that can be turned into the problem-mechanism-species schema.
Relevance to the query problem during data generation is preserved through contextualized prompts that integrate constraints and problem descriptions, which are parameterized in the current system implementation and is straightforward to adapt to new user query.
At run time, active features also maintain relevance through contextualized prompts, which are also currently parameterized in the prompt design for adaptation to user input.}

\vspace{-1em}
\subsection{Recognition}
\subsubsection{Recursive Clustering of Active Ingredients} \label{subsubsection:new_backend_recursive_clustering}
One way to help users recognize relevant inspirations is to cluster organisms with similar mechanisms together to improve the efficiency of scanning for diverse inspirations.
In order to organize the active ingredients in semantically meaningful groups, we adapted a density-based algorithm \codett{DBSCAN} to re-cluster those mechanisms it could not initially cluster with a gradually relaxing minimum distance parameter $\epsilon$.
This approach had the benefit of fixing the relatively straightforward clusters (\ie groups of mechanisms with very similar surface text forms, resulting in very low distances among them in the embedding space) early on, and by lowering the sensitivity threshold subsequently in order to identify less obvious yet coherent clusters (\ie groups of mechanisms that look different in the surface text form yet are semantically related, resulting in relatively higher distances among them).
After the final run of the algorithm, mechanisms that could still not be clustered were appended to the end of the list.
\subsubsection{Active Ingredient Extraction}
\label{subsubsection:new_backend_active_ingredient_extraction}
Another way to improve recognition brought up by participants in our pilot study is to generate cluster descriptions more informative about their \textit{active ingredients}, or transferable concepts that enable the mechanisms.
To do so we used GPT-4 (\codett{gpt-4-turbo-preview}) with a prompt (Appendix~\ref{appendix:active_ingredient_extraction}) to extract active ingredients from mechanism outputs from the earlier diversification and goal-driven generation steps (\S\ref{subsection:data_discovery_diversification_structure}).
In the system message we instruct the following  for identifying active ingredients that we found useful from pilot testing: 1) shorter length (\ie 15 words or less), which was easier to skim and increased the cluster separation by excluding secondary features of commonality among the species, 2) descriptions with a verb or a verb phrase, which was easier to parse as they often presented the information in the form of `what acts upon what', and 3) concrete active ingredient examples.
In the user message we provide the mechanism description to process accordingly.

\vspace{-1em}
\subsection{Transfer}
\subsubsection{Sparks} \label{subsubsection:sparks_generation}
As discussed in prior work, it is cognitively demanding to transfer an inspiration to the target domain of a design problem, involving identifying the relevant features to transfer, how those features map to features in the new domain, and adapting any unmapped or missing elements~\cite{holyoak1989analogical, gentner1983structure}.
There is also a question of how and when such a mapping should take place -- it is expensive both computationally and from a user-attention perspective to compute and show mappings for every potential inspiration.

Here we introduce the interaction paradigm of `sparks' to support analogical transfer and the computational approach to enabling them.
Sparks are generated when a user interacts with an inspiration to save it as potentially interesting, passing an initial threshold of interest but not necessarily requiring the user to have deeply engaged with mapping that inspiration to how it would work in the target domain, \ie leveraging a `spark' of interest from the user.
Within seconds of indicating interest in an inspiration, the system generates two spark cards at the top of the sidebar stream, attracting the user's attention.
Text on the spark cards is large enough for the user to start reading the first few lines with little effort, helping them to see how the inspiration could be applied to the target domain.
The generation of two sparks aims to prevent fixation in the use of the inspiration and present multiple solution paths which could lead to schema induction and the user exploring even more of the design space.
Thus, the design of sparks is aimed at scaffolding the mental process of transfer by reducing cognitive effort at each step and directing attention to potentially fruitful areas in an approach aimed at leveraging natural practices and that recognizes the designer's need for agency.

Two sparks are generated using GPT-4 (\codett{gpt-4-turbo-preview}) each time the user saves an inspiration or clicks on the `spark' button (fig.~\ref{fig:interface}, \cirnum{D}) on a mechanism.
Each spark card also includes helpful features such as a caret for expanding/collapsing the card, the timestamp of creation, a clickable thumbnail (fig.~\ref{fig:interface}, \cirnum{J}) showing the source mechanism, which expands the modal view upon clicking it, and control buttons (fig.~\ref{fig:interface}, \cirnum{K}) for further generating new sparks of the spark content, Q\&A, and deletion. The content of each spark is directly editable.
The spark-generation prompt (Appendix~\ref{appendix:spark_generation}) contextualizes the user-selected mechanism inspiration with the design problem description and the constraints provided with the problem.
We instruct GPT-4 to be succinct when generating a spark (\ie under 500 characters) and to provide a descriptive title.

One challenge when developing sparks was that generating multiple sparks for the same mechanism inspiration led to highly similar sparks, despite explicit instructions included in the prompt that requested diversification in generation.
To address this, we adapted feedback-augmentation approaches such as self-refine~\cite{madaan2024self} to improve diversity by adding the most recently generated 20 sparks as part of the prompt, and requesting that the new generation be novel and not redundant with them.
Through validation on 3800 pairs of sparks we found that semantic diversity (as measured by cosine similarity, \cf~\cite{gero2022sparks,hayati2023far,tevet2020evaluating}) was significantly higher when precedent-based diversification was used (M=.24, SD=.073) than not (M=.17, SD=.090) (\tind{7291.87}{-42.41}{<< .0001}). See Appendix~\ref{appendix:precedent_based_diversification} for more analysis details.
\vspace{-1em}
\subsection{Elaboration}
\subsubsection{Trade-off Analysis} \label{subsubsection:trade-off}
As discussed in both the workshop and the design probe as well as in prior work, the process of analogical transfer in design rarely ends with mapping.
Designers often need to elaborate and drill down on the particular mechanisms suggested by the inspirations to understand how they would actually work, their physical limits, and pros and cons of different approaches in the context of the design problem.
To address this with a consistent interaction paradigm as the spark cards we introduced `trade-off' cards.
The goal of trade-off cards is to not only help users who are looking for specific elaborations already, but to proactively suggest dimensions on which users might want to elaborate and drill down further.

We generate a new trade-off analysis card using GPT-4 (\codett{gpt-4-turbo-preview}) each time the user clicks on the trade-off button (fig.~\ref{fig:interface}, \cirnum{E}) on a mechanism.
We design a trade-off analysis prompt (detailed in Appendix~\ref{appendix:trade_off_generation}) to request the generation. 
In the prompt we contextualize the user-selected mechanism inspiration using the design problem description and the constraints provided with the problem.
We instruct GPT-4 to return the `pros' and `cons' of the mechanism inspiration in the context of the design problem using a markdown table format that places each pro-and-con pair in a new row, and give each item in the table a succinct (3 words or less) label.
In the view, we display the analysis in each trade-off card in the stream (fig.~\ref{fig:interface}, \cirnum{I}) and implement a scrollable and formatted table view using \codett{React-Markdown}\footnote{\url{https://github.com/remarkjs/react-markdown}} and \codett{remark-gfm}\footnote{\url{https://github.com/remarkjs/remark-gfm}}.

\subsubsection{Q\&A Handling}
\label{subsubsection:q_and_a}

Many of the elaboration questions brought up in the pilot study by participants, such as ``I wish I could know more about the lubrication mechanism of slimes'', are not easily captured in preset queries such as trade-offs.
To support more flexible elaboration we introduce Q\&A cards, which allow the user to ask any question they would like to the LLM with the context of the design problem and inspiration already included.
Q\&A cards also serve as a general purpose catch-all to probe what kinds of questions users might want to ask that haven't yet been explicitly designed for in the system.

\revised{}{To fluidly respond to free-form user questions in the Q\&A text area (fig.~\ref{fig:interface}, \cirnum{F}), we send GPT-4 a prompt that instructs the model to answer the question and includes information about the source mechanism and the design brief context in its system prompt portion, while the user prompt portion includes the question itself (fig.~\ref{fig:q_and_a_prompt}, Appendix~\ref{appendix:q_and_a}).}
\revised{}{To help users understand and recall the provenance of the generated response, we send another GPT-4 request for a rationale and appropriateness assessment. This rationale is then featured as a tooltip next to the timestamp on the response card (fig.~\ref{fig:interface}, \aptLtoX[graphic=no,type=html]{\aptbox{?}}{\cirnum{?}} icon in the header of each card in the stream \cirnum{I}).}

\subsubsection{Drill-down on related research} 
Another aspect of drilling down on an inspiration raised in the pilot study was a request for more information about a particular mechanism, specifically web or scientific resources about it.
Although not a primary focus of our main study goals, we did include a rudimentary support for this need to probe users' interest, through linking them to a generative AI search engine that provided cited web sources for a given mechanism (details of its implementation in Appendix~\ref{appendix:subsection_perplexity}.

\section{User Study}
To investigate how \sys{} affected bio-inspired design generation we conducted a within-subjects study in which participants experienced both \sys{} and a baseline condition for two design problems.
Our goal in this study was to evaluate the combined value of the design decisions made in \sys{} on creativity-related outcomes including the quality, amount, and diversity of ideas generated, while using an analysis of participant think-aloud and post-study interview and survey data to explore participants' experiences with different features and functionality.
Towards this goal we explored appropriate baseline conditions for comparison that would provide users with a similar set of resources while corresponding to what designers in our workshop and pilot study might access for bio-inspired design today.
The final baseline condition selected provided participants with access to AskNature.org, a popular bio-inspired design inspiration site which served as the seeds to \sys{}'s generation algorithm, and with a ChatGPT terminal.
We leave for future work more nuanced conditions such as ablation studies of various features or augmentation of the ChatGPT terminal (such as through persistent context, custom GPTs, or fine-tuning).

\vspace{-1em}
\subsection{Methodology}

We employed a within-subjects study design to compare \sys{} with a baseline system for inspiration and a shared Google Spreadsheet participants accessed to write down their own ideas.
We chose two design problems for user ideation, including how to design wheelchairs that allow users to go up the stairs easily and how to design an innovative bike rack for sedans.
These problems were chosen because they involve multiple, potentially competing constraints (\eg lightweight but durable) and were pilot tested for being able to be completed within the timed ideation task. 
\begin{lstlisting}[style=prompt]
(The 'Wheelchair' problem) Design wheelchairs that can also allow users to go up the stairs easily.
Constraint 1 (Lightweight yet Durable Construction): The wheelchair should be lightweight and be able to withstand a heavy load without structural failure.

Constraint 2 (Compact and Foldable Design): The wheelchair must be foldable to a 1/4 of the volume within 30 seconds without the use of tools.
\end{lstlisting}

\begin{lstlisting}[style=prompt]
(The 'Bike rack' problem) Design innovative bike racks for sedans.
Constraint 1 (Integration without Interfering with Aerodynamics): The bike rack's design must not significantly reduce the vehicle's fuel efficiency when installed and with bikes mounted.

Constraint 2 (Versatility in Accommodating Different Bike Types): The rack must be able to securely hold at least three different bike frame sizes (e.g., 16", 20", and 26") without the need for additional adapters.
\end{lstlisting}
We randomly assigned problems to conditions for the main timed tasks (20 minutes each), counterbalancing the order of presentation using three 2x2 Latin Square blocks.
Participants followed a fixed procedure in the study, which took place remotely using Zoom: Introduction, Consent, Demographics survey; Tutorial (detailed in Appendix~\ref{appendix:system_tutorials}) of the first system via screensharing; Main task for the first system (20 min); Rating task for the first system (only in the \sys{} condition); Survey for the first system; alternating and repeating for the second system; followed by a debrief.
Participants were asked to share their screen during the timed tasks and think-aloud.

To probe how participants felt about the utility of different information generated using various AI-based system features, after \sys's main task, participants were also presented with a rating interface that showed a list of saved Sparks, Q\&A, and Trade-off cards along with a 5-point Likert-scale for them to rate their usefulness in their process.

To analyze perceptions of usefulness we collected participants' subjective ratings to a modified Technology Acceptance Model survey questionnaire items focused around task performance and easiness of learning from~\cite{tam_survey} (4 items) using a 7-point Likert scale (1: \textit{strongly disagree}, 7: \textit{strongly agree}).
In addition, we employed questions focused on serendipity and exploration adapted from~\cite{niu2018surprise,mccay2011measuring} (9 items) and the questions on the value of AI assistance and the quality of inspirations found in the system.

\subsubsection{Participants}
We recruited 12 researchers (7 women, 5 men) through advertisement on Upwork\footnote{\url{https://www.upwork.com/}} and email lists at a State Arts College.
Participants' background included professional UX design experience (6), professional illustration and graphic design (1), PhD in Psychology (3), and a current undergraduate student in Arts and Design (1) and a master's student in AI and Data Science (1).
Participants' average age was 36.1 years (SD=9.91).

\subsubsection{Baseline}
Participants were given 5 URLs from AskNature, each pointing to a functional category equivalent to those that were used for the \sys{} backend dataset pipeline: Manage Impact, Manage Tension(Manage Tension), Manage Compression, Manage Turbulence, and Modify Speed\footnote{(Manage Impact) \url{https://rb.gy/rvz17u}; (Manage Tension) \url{https://rb.gy/t3se2z}; (Manage Compression) \url{https://rb.gy/xvogjb}; (Manage Turbulence) \url{https://rb.gy/9apgoq}; (Modify Speed) \url{https://rb.gy/r7o2c8}}.
Before the baseline task began, participants organized their screen by opening up all 5 tabs in their browser on the left-hand side of the screen and sign-in and open the ChatGPT\footnote{\url{https://chat.openai.com/}} interface on the right-hand side of the screen.
They were instructed to freely use the platforms to help themselves understand and ideate with mechanism inspirations found on AskNature for the design problems.
Each participant was also instructed to write down the ideas they come up with in the process in a prepared Google spreadsheet, with a brief description of the species that inspired each idea.

\section{Results}
\subsection{Creative idea quality and quantity}
We begin discussion of the results with one of the most critical questions: did \sys{} result in higher quality creative ideas? Although creative quality can be defined and operationalized in many ways, one common approach in the creativity literature is to break down an idea's creative quality into the dimensions of novelty, value, and feasibility~\cite{shah2003metrics, yu_analogical_ideation_chi14}.
By combining these three dimensions, often using a penalizing function for being low on any dimension such as the geometric mean, researchers have attempted to balance trade-offs between these factors into a holistic judgment of creative quality.

Following this approach we created a rubric operationalizing novelty as how unique versus common an idea is found to be in the world.
Value corresponded to how effectively the idea addresses the main challenge in the problem (\eg how well does the idea directly help wheelchair users go up the stairs more easily?) as well as the main constraints (\eg how lightweight or foldable is it?).
Feasibility was operationalized as the ease of achieving the creation of the idea with current resources and technology.
These three dimensions were combined using the \textit{geometric mean} into an aggregated score representing the overall creative quality of an idea.

To judge quality we recruited as a domain expert a senior PhD student at an R1 institute in North America with expertise in \rev{building systems augmenting} wheelchair users\rev{' mobility. The expert recruited had practical experience in designing products employed by wheelchair users, including broad knowledge of the feasibility of both physical constraints such as the materials used and their configuration and reliability, as well as user constraints such as what types of products would be likely to be used by wheelchair users. }The first author and the expert met to discuss the assessment rubric using 5 randomly selected ideas for each of the two design problems used in the study \rev{(see {Appendix~\ref{appendix:judge_scores_rationale_samples}})}.
Upon reaching agreement on the rubric, the first author and the expert independently judged the quality of remaining ideas blind to condition.
Intraclass Correlation Coefficients showed a significant level of reliability between the judges on the independently coded set of examples: 0.94 (Novelty), 0.74 (Feasibility), 0.90 (Value), following ~\cite{yu_analogical_ideation_chi14} in which the more conservative ICC(2,k) method is used.
\rev{To further validate the reliability of the feasibility scores, which could be especially sensitive to domain expertise, we also involved an experienced product design expert (one of the paper's authors) who independently assessed each idea blind to condition and provided detailed rationales for the same 10 examples the two judges used for discussing the rubric. 
This expert had an undergraduate degree in engineering design and a Master's and Ph.D. in Mechanical Engineering specializing in engineering design, with over 16 years of experience in developing mechanical, mechatronic, and software-based systems. An analysis of their reliability scores showed a strong and significant correlation with the judges' scores (Pearson's correlation coefficient = 0.85, $p = .002$).}

The final scores were computed by averaging the two judges' scores \rev{on each dimension and combining them using the geometric mean into a single overall score, as per common procedure in the literature (\eg~\cite{shah2003metrics,yu_analogical_ideation_chi14}). 
The summary statistics of scores in each dimension are as follows:
\begin{table}[ht]
\centering
\begin{tabular}{lccc}
\hline
\textbf{Condition} & \textbf{Novelty} & \textbf{Feasibility} & \textbf{Value} \\
\hline
Baseline & $3.9$ ($1.67$) & $6.8$ ($1.37$) & $3.8$ ($1.68$) \\
\sys{}   & $7.1$ ($1.03$) & $5.7$ ($1.02$) & $7.0$ ($0.91$) \\
\hline
\end{tabular}
\caption{Means with standard deviations in parentheses for the three dimensions (Novelty, Feasibility, Value) across the two conditions (Baseline, \sys{}).}
\vspace{-2.5em}
\label{tab:mean_sd}
\end{table}}

We analyzed overall scores using a linear mixed-effects model to take into account potential participant-specific effects of our within-study design.
An initial model compared the control and experimental conditions (\ie Baseline vs. BioSpark) with participant as a random effect, using the Residual Maximum Likelihood (REML) method for fitting model parameters~\cite{muradoglu2023mixed}. We found that while the residuals were normally distributed (through a visual examination of the Q-Q plot of residuals showing good alignment to the line of estimation and the complementing result of the Shapiro-Wilk test showing residuals not significantly deviating from normal distribution, $p = .06$), the homoscedasticity assumption was likely violated from the results of both Breusch-Pagan and White's tests ($p < .001$ in both cases). We thus modified the model to use a covariance estimator adjusting for variance heteroscedasticity and residual correlation, and find that this final model showed a good fit to the data ($F = 802.87, p < .001$) overall, with an $R^2 = .58$ indicating that approximately 58\% of the total variability in scores was explained by the model.

\begin{figure}[ht]
        \centering
        \begin{tabular}{lcccccc}
        \hline
         & \textbf{Coef.} & \textbf{Std. Err.} & $t$ & $P>|t|$ & \textbf{[0.025} & \textbf{0.975]} \\
        \hline
        \textbf{Const.} & 4.4802 & 0.102 & 44.105 & 0.000 & 4.280 & 4.681 \\
        \textbf{Cond.} & 2.0396 & 0.126 & 16.221 & 0.000 & 1.792 & 2.288 \\
        \hline
        \end{tabular}
        \vspace{-1em}
        \caption{Regression results show that the estimated overall quality for the baseline condition is 4.48 and 4.48 + 2.04 = 6.52 for the \sys{} condition. The effect of condition was significant, $t = 16.22$, $p < .0001$, suggesting a substantial difference between the baseline and the BioSpark condition.}
        \label{tab:regression_results}
    \end{figure}%

\begin{figure}[ht]
        \centering
        \includegraphics[height=4.5cm]{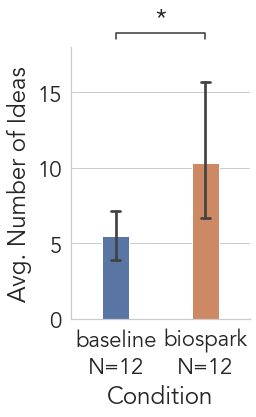}
        \vspace{-1.3em}
        \caption{The number of participants' ideas during the experiment.}
        \label{fig:idea_quantity}
    \vspace{-1.5em}
\end{figure}

Results showed that the overall creative quality estimate was significantly higher in \sys{} (M=6.5, Standard Error=.16, $t = 66.62$, $p < .001$) than in the Baseline condition (M=4.5, Standard Error=.10,  $t = 28.34$, $p < .001$) (Table~\ref{tab:regression_results}).
Representative ideas from each condition are shown in Table~\ref{tab:representative_ideas} to illustrate how high vs. low quality ideas differed for each design problem.
\begin{table*}[ht]
    \centering
    \renewcommand{\arraystretch}{1.2}
    \begin{tabular*}{\linewidth}{@{\extracolsep{\fill}}clcccp{.5\linewidth}}
    \hline
    \textbf{Rank} & \textbf{Cond.} & \textbf{Prob.} & \textbf{Score} & \textbf{An-Tr?} & \textbf{Idea} (Summarized) \\
    \hline
    Top & \sys{} & 2 & 7.86 & Y & {\small This bike rack uses a \textbf{bio-mimetic spring mechanism}, inspired from \textbf{Anura's powerful legs}, to stretch (and compress) to accommodate different bike sizes. Its skeletal structure, mimicking frog bones, flexes to absorb road vibrations, protecting bikes. Aerodynamically shaped to reduce drag, it `leaps' into a compact form when not in use, preserving fuel efficiency.}\\
    Bottom & \sys{} & 1 & 3.98 & Y & {\small \textbf{Retractable `legs'}; Jointed segments that extend and contract to climb.}\\
    \hline
    Top & Baseline & 2 & 6.80 & Y & {\small \textbf{Aerodynamically shaped bikerack} design inspired by how \textbf{marine mammals like dolphins and whales} have evolved flippers and tail fins that optimize their movement in water. Their body shape tends to be more rounded than that of fish but is streamlined for efficient travel. The tail fins (flukes) provide powerful propulsion, while the pectoral flippers are used for steering and stabilization.}\\
    Bottom & Baseline & 1 & 2.62 & N & {\small Golden bamboo for durability and lightweight design.}\\
    \hline
    \end{tabular*}
    \caption{Top-1 and Bottom-1 scoring ideas from each condition, and whether they perform analogical transfer. `Cond.' represents the condition in which the idea was produced; `Prob.' represents the problem the idea is for (1: the `wheelchair' problem; 2: the `bike rack' problem); `Score' represents the geometric mean of expert-judged novelty, value, and feasibility scores; `An-Tr?' is a binary value representing the presence of Analogical Transfer; The Bottom-1 idea is summarized for conciseness. Boldfaced text in the idea description represents the source and the target in analogical transfer. Full table in Table~\ref{tab:representative_ideas} (Appendix~\ref{appendix:top5_bottom5_ideas}).}
    \vspace{-3em}
    \label{tab:representative_ideas}
\end{table*}
High quality ideas included several examples of analogical transfer, as shown in an idea inspired from tree frogs' adhesive toe pads to harness the rear surface of the vehicle itself for securing bikes (Top 5th idea in \sys{}) or an idea inspired from sea anemones' burrowing to construct a contracting and expanding wheelchair base (Top 50th idea in \sys{}).

Interestingly, there was no evidence of a trade-off of idea quality with idea quantity, with \sys{} users generating nearly twice the number of ideas (M=10.3, SD=8.46) than in the baseline condition (M=5.5, SD=2.91), which a paired two-tailed t-test suggests was a significant difference (\tpaired{13.56}{-2.35}{=.04}) (fig.~\ref{fig:idea_quantity}). Generating more ideas, particularly diverse ones, is itself a design goal for many designers; thus, the significant increase in the number of ideas in combination with the large increases in the novelty and value dimensions suggests a promising direction.

In summary, from several converging angles our results suggest that \sys{} led to participants generating more and higher quality ideas than in the baseline condition.
Participants' own perceptions of the value of the system appeared consistent with this, with participants' agreement with the statement `\textit{Using this system would improve my task performance}' was significantly higher in the \sys{} condition (M=6.5, SD=.80,\includegraphics[scale=0.3]{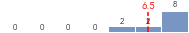}) than in the baseline condition (M=5.3, SD=1.07,\includegraphics[scale=0.3]{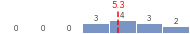}) (\wilcoxon{2.0}{.02}).

\subsection{Diversity} 
Another of our design goals was to not only help users engage with inspirations, but for those inspirations to be diverse and help them explore new design spaces they might not have thought of on their own. The previous analysis on quality is consistent with the hypothesis that participants' increased novelty in idea generation may have been based on encountering more diverse inspirations; in this section we investigate the role of diversity more directly.

In the context of bio-inspired design, one measure of diversity in design space exploration is how many different species in nature  participants are engaging with for ideation, which has potential for not only inspiring ideas based on a specific mechanism of the particular species but also opening up a new space of design that encompasses other mechanisms of the species or its related species.
To quantify the number of unique species mentioned by a participant we extracted the inspiring species' names using \codett{gpt-4-turbo-preview} with a prompt (Appendix~\ref{appendix:species_extracton}).
A paired two-tailed t-test revealed that the number of unique species used for inspiration was significantly higher (nearly double) in the \sys{} condition (M=8.2, SD=4.97) than the baseline condition (M=4.6, SD=2.71) (\tpaired{17.02}{-3.30}{=.007}).
Although it is difficult to directly quantify the effects of using more species, a plausible hypothesis is that using more species provided a greater number of mechanisms for participants to leverage in solving the design problems, resulting in both more ideas as well as those ideas being higher value and more novel as described in the section on quality.

The interview, survey, and observation data support a deeper but consistent understanding of which aspects of the \sys{} design and interactive features most contributed to broadening participants' exploration of design spaces.
Participants felt that the AI features in \sys{} helped them be more creative.
As will be described in the Feature Use (\S\ref{subsubsection:sparks}) section below, participants viewed the `spark' generation feature as usefully nudging them to reframe the problem (P6), creatively adapt and translate the source mechanism inspirations into the target domain (P12), beyond just providing useful inspirations or ideas (P1, P7).
Consistent with these statements, participants' agreement with the statement `\textit{I was able to examine a variety of inspirations}' was significantly higher in the \sys{} condition (M=6.6, SD=.67, \includegraphics[scale=0.3]{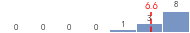}) than the baseline condition (M=5.4, SD=1.62,\includegraphics[scale=0.3]{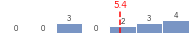}) (\wilcoxon{0.0}{.03}).

In contrast, the baseline condition felt less dynamic and led to less exploration by participants.
For example, while P4 thought the functional category-based organization in AskNature helped his navigation, it also somewhat fixed the broader design space he explored in, as he felt like \pquote{got stuck somewhere a little bit because I came up with this sideways top-mount bike rack idea early on from the `manage turbulence' concept} (P4).
He also thought it was easier to see the relevance of mechanisms in \sys{} and follow-up with more exploration.
However, it is possible that such structured organization could be profitably combined with a more dynamic system, as described by P9 that he is \pquote{not as interested as working from the ``bottom up'' for research, and would instead like to have AI help brainstorm from the ``top down''} (such as through the functional organization in AskNature). 

\subsection{Feature use} \label{subsection:feature_use}
\begin{table*}[h!]
\centering
\vspace{-1.1em}
\begin{tabular}{p{2.2cm}|p{3cm}|p{5.5cm}}
\hline
\textbf{Feature Name} & \textbf{Usage Freq. (M, SD)} & \textbf{Usefulness (M $\downarrow$, SD)} \\ \hline
Stream & N/A & 6.3 (SD=0.98) \includegraphics[scale=0.3]{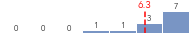}\\ \hline
Spark & 6.2 (SD=3.90) & 5.9 (SD=1.08) \includegraphics[scale=0.3]{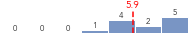} \\ \hline
Q\&A & 2.3 (SD=1.97) & 5.8 (SD=1.19) \includegraphics[scale=0.3]{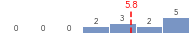} \\ \hline
Trade-off & 1.3 (SD=1.66) & 5.2 (SD=0.94) \includegraphics[scale=0.3]{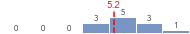} \\ \hline
Clustering & N/A & 4.1 (SD=1.44) \includegraphics[scale=0.3]{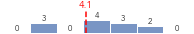} \\ \hline
\end{tabular}
\caption{Usage and usefulness ratings for \sys{} active features, measured on a 7-point Likert scale (higher is better).}
\vspace{-2.5em}
\label{tab:feature_usage}
\end{table*}
The above analyses suggest that the design choices made in \sys{} led to higher quality ideas, more ideas, and consideration of a more diverse inspiration space.
In this section we aim to investigate which features were used by participants and their perceptions in order to better understand the results above and to unpack which design choices were more successful than others.
In the below sections we characterize through think-aloud, usage, and survey data participants' experience with various system features.

\subsubsection{The Stream Interface}
The stream interface design of \sys{} was aimed at providing a scratchpad and organization space for thought, aiming to help users triage and interact with AI support. The stream was perceived as helpful for participants during ideation and it was the highest-scoring system design feature in \sys{} (M=6.3, SD=.98; Table~\ref{tab:feature_usage}).
On its value, P1 said:
\begin{quote}
\pquote{I liked these because they kept my thoughts and all the information very organized. It allowed me to focus on the actual text vs focusing on the organization of everything. It would have been even more helpful though if there was a way to enable bullet points or formatting tools within these. I would have used bold or italics for example.}
\end{quote}
P4 noted that \pquote{It's great to see all the ideas in one place. It provided a nice anchor to go back to. It's also a nice touch to have a trash bin so that I can go back and check the ideas I discarded earlier.} and P5 said the design was helpful for \pquote{note-taking and brainstorming through relevant results and queries}.

Participants also commented on how the design and presentation of information in \sys{} streamlined their exploration and helped them accomplish the task.
P10 described it as: \pquote{I like that it's integrated into one space. I can press a button to get to the particular need that I had.}, and P7 mentioned that \pquote{[the stream organization of information] helped me to compare and contrast} while P10 commented \pquote{[the stream was] definitely helpful. [It] allowed me to narrow down best options.}.

These comments are consistent with participants' agreement with the statement `\textit{I could easily explore many inspirations without getting lost}' (M=6.3, SD=.89 in \sys{},\includegraphics[scale=0.3]{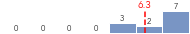}, M=5.2, SD=1.80 in baseline,\includegraphics[scale=0.3]{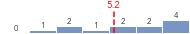}, \wilcoxon{2.0}{.05}).

\subsubsection{Mechanism Clusters}
Next, we investigate how active ingredient clustering, designed to help improve participants' recognition and efficiency of scanning diverse mechanism inspirations might have helped.
Several participants mentioned clustering as helpful for focus, navigation, and comparison, as it \pquote{helped to categorize the needs/constraint which I am focusing on \& pick the one which align well with my end goal.} (P6); \pquote{navigating the problems and finding solutions to associated animals and their unique features} (P8), and \pquote{giving me information to help drive a decision and look at alternative options.} (P5).

The clustering structure appeared to be a functional entry point for exploration even for those who did not paid a lot of attention to: \pquote{I found myself not exactly focusing on the ``clustering'' per say, but I did focus on the specific mechanisms that seemed could be a good way to enhance design features.} (P1). However, participants appeared to use the drill-down functionality into clusters less, either not attending to the species that would show up when drilling down into a particular cluster or even being unaware that the drill-down existed: \pquote{I didn't look into the other species within the clustering so much as the responses from the AI} (P11); \pquote{didn't use it, was busy using other tools} (P2); \pquote{I actually didn't notice the clustering} (P3). 

\subsubsection{Sparks}
\label{subsubsection:sparks}
The `Spark' button was the highest-used feature among the active features, at 6.2 times (SD=3.90),
and was generally perceived as useful (M=5.9, SD=1.08). P1 described its value as:
\begin{quote}
\pquote{The ``Sparks'' button provided me with a lot of interesting insights and got me thinking in directions I may not have thought of on my own. It spurred my creative thought.}
\end{quote}
Other participants mentioned that sparks \pquote{provided more inspiration/ideas} (P7), \pquote{helped me with better viewpoint and perception which I normally wouldn't think of} (P6), \pquote{makes me think of a new design space} (P12), and \pquote {gave me inspirations on how I could translate the idea into design criteria and functions I want to achieve} (P12).
The `Spark' button was also often used in combination with the Q\&A feature, which provided users opportunities to follow-up on initial sparks to more deeply understand and further develop them (see Table~\ref{tab:intent_query} for different query intent users submitted in the `Q\&A' text box).

Reflecting these observations, We find a marginally significant difference in the agreement levels with the statement:
`\textit{The system enabled me to make connections between different inspirations}' showed a marginal difference between \sys{} (M=6.5, SD=.80, \includegraphics[scale=0.3]{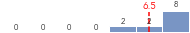}) and Baseline (M=5.6, SD=1.56,\includegraphics[scale=0.3]{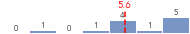}), (\wilcoxon{1.5}{.07}).

\subsubsection{Trade-offs} \label{subsubsection:trade_offs}
Participants found the trade-off cards generally useful (M=5.2, SD=.94) and mentioned how it helped them to prioritize and triage ideas for further consideration. 
As P6 said: \pquote{It very much helped me to decide which feature/constraint I am willing to give up or trade off. It showed me valid reasons to pick the side correctly.} (P6). In this vein P10 also mentioned \pquote{It was good to see the pro and cons of a concept to see which were stronger to follow up with.} (P10).

Interestingly, several participants mentioned they noted the cons in the pros/cons analysis as more important than the pros list. P11 and P12 described the reason as: \pquote{I tend to look at cons first and it directs and warns me what are the constraints I'll be working with} (P12); \pquote{This Pros and cons is helpful, to look at the weight concerns. When I look at pros and cons usually I go straight to the cons, the reason, is personally I focus on bad things first.} (P11).

However, there were also limitations of the trade-off analysis, namely how the dimensions of comparison was not readily configurable by the user, which sometimes led to useful aspects of comparison but other times not. On this point, P1 described: \pquote{I used it for moreso looking for cons of the mechanism that could act as barriers for the design itself, but the button provided me insights that were moreso focused on the cons being related to business expenditure and manufacturing.} (P1).
P4 mentioned having to save ideas that use different materials and click the trade-off button on each even when he wanted a direct comparison on the material and manufacturing cost aspects across multiple ideas: \pquote{interesting, but it would be more helpful if I could ask about the pros and cons of a material or mechanism } (P4). Similarly, P5 said that \pquote{I want it to be able to compare this spark and this other spark in terms of strengths and weaknesses} (P5).

\subsubsection{Q\&A cards} \label{subsubsection:q_and_a}
The Q\&A cards introduced in \sys{} aimed to help flexible user interaction with interesting mechanism inspirations, such as elaborating on relevant aspects of the design problem or gaining a deeper understanding of mechanism inspirations they want to build on.
Participants thought the `Q\&A' feature generally useful (M=5.8, SD=1.19) and mentioned that it \pquote{helped clarify the details of the general overview of the idea} (P3), and that \pquote{I could ask some very specific questions about very specific mechanisms such as finding a fabricated material that is comparable to chitin, and get a useful reply} (P4).
P6 also commented how the AI assistance \pquote{allowed for interactive sessions to implement my ideas along} (P6).

The intent of user queries within the Q\&A interaction seemed to have varied, characterized as the following four types (Table~\ref{tab:intent_query}): 1) (\textbf{Understanding}) to more deeply understand design ideas' working and relevant physics, 2) (\textbf{Adaptation}) to creatively adapt the ideas into new directions, 3) (\textbf{Constraints}) to probe how they would meet certain constraints to become value, and 4) (\textbf{Material}) to engage in material selection and engineering feasibility.
In our observations, participants seemed to often engage in an initial deeper understanding of an interesting design inspiration (Understanding), followed by adapting them further by generating adaptation ideas (Adaptation), and increasing their value by considering how various design constraints would be addressed (Constraints) and how engineering and manufacturing feasibility might be enhanced (Material).
Table~\ref{tab:intent_query} shows representative user queries of each type.
\begin{table*}[h]
\centering
\begin{tabular}{p{0.15\textwidth} p{0.8\textwidth}}
\toprule
\textbf{Intent} & \textbf{Sample Q\&A Query}\\
\hline
\multirow{2}{*}{Understanding} & \textit{``How would this be able to collapse the frame itself, not just for the bikes while on the rack?''} \\
 & \textit{``Tell me more about how this design ensures minimal air resistance, maintaining fuel efficiency.''} \\
\hline
\multirow{2}{*}{Adaptation} & \textit{``How to modify this in a mobile space?''} \\
& \textit{``is there a way to install modular adapters? ''}\\
\hline
\multirow{3}{*}{Material} & \textit{``Material wise what are some of the examples that could withstand a similar force but in a large scale?''} \\
 & \textit{``What kind of materials would I need to make this hydrophobic?''} \\
\hline
\multirow{2}{*}{Constraints} & \textit{``How does a user without functioning limbs use this wheelchair?''} \\
 & \textit{``Can this bike rack safely restrain e-bikes?''} \\
\bottomrule
\end{tabular}
\caption{Intents and Corresponding User Queries}
\label{tab:intent_query}
\end{table*}

\subsection{Engagement beyond Inspiration} 
One of the main goals of \sys{} was to support users beyond finding inspirations, to helping users notice the relevance of,  transfer, consider tradeoffs, and elaborate on inspiration mechanisms in the domain of the target design problem.
Engagement is particularly relevant to analogical inspirations that may require significant cognitive effort for valuable transfer to occur ~\cite{gentner1985analogical,gick1983schema}. 
We explore this research question using the interaction log data as well as transcribed participants' think aloud data.

\subsubsection{Interaction Log Data Analysis Results} Participants in the \sys{} condition demonstrated active engagement with the inspiration mechanisms.
On average, they generated 18.8 sparks (SD = 14.28), wrote 1.0 ideas from scratch (SD = 1.15), and deleted 9.4 sparks (47\%; SD = 8.74) after evaluating their relevance or lack thereof.

Furthermore, as detailed in \S\ref{subsection:feature_use}, participants found the Q\&A feature useful, engaging with it an average of 2.3 times (SD=1.97) to adapt inspiration mechanisms and enhance their understanding.
Their range of queries and intent is summarized in Table~\ref{tab:intent_query}.
These findings collectively suggest that participants actively engaged with \sys{}-generated sparks, leveraging them to notice relevance, transfer ideas, and elaborate on inspiration mechanisms.

\subsubsection{Think-aloud data analysis results}
To explore the research question further we analyzed a subset of the transcribed participants' think-aloud along with their behavior descriptions.
The research team met to discuss coding of interview and think-aloud data from the study, with one salient feature of the data being that participants seemed to engage with mechanism inspirations differently in depth, for example with or without follow-up actions that related to attempting to deepen their understanding of the inspirations, of their relevance to the design problem and of trade-offs regarding different design constraints, and attempting to come up with new ideas that could adapt the inspirations to a design problem in new ways.

In order to characterize these differences the first two authors developed four codes for participants' different engagement patterns corresponding to two common types of shallow engagement and two common types of deeper engagement:
\begin{itemize}
    \setlength{\itemindent}{-2em}
    \item[]{[\textbf{S1}: ``\textit{Interesting!}'']}: Positive comments on a mechanism inspiration, but directly followed by moving on to a different mechanism that was visible to the participant.
    \item[]{[\textbf{S2}: ``\textit{I'm not sure how this might be relevant}'']}: Negative comments on a mechanism inspiration, but similarly followed by moving on to a different mechanism  that was visible to the participant.
    \item[]{[\textbf{D1}: \textit{Engaging with relevance understanding and constraints consideration}]}: Engagement with AI to understand a mechanism inspiration's relevance to the design problem, for example by asking the following types of questions ``\textit{tell me examples of...}'' or ``\textit{how might this be used/applied...}''
    \item[]{[\textbf{D2}: \textit{Actively coming up with new ideas}]}: Exploring the design space and actively generating new ideas ``\textit{it made me think of...}''
\end{itemize}

One author transcribed the interview and think-aloud data from the study, and incorporated descriptions of participants' actions with each platform (\eg what the participant is typing in the ChatGPT interface or what the participant is clicking in \sys), that participants' think-aloud did not describe but were relevant to understanding their engagement process and intent.
This resulted in 266 transcripts across 12 participants.
Coders coded a set of randomly selected 16 transcripts together blind to condition and built consensus through discussion.
They then coded 30 additional randomly selected transcripts independently.
The inter-rater agreement of codes for this set showed a moderate to strong level of agreement $\kappa = 0.76$.
Thus, the first author coded the remaining 218 transcripts alone.
Table~\ref{table:ex_actions_and_codes} shows representative cases and how they map to the four codes of engagement patterns.

\begin{table*}[ht]
\centering
\begin{tabular}{c p{2.25cm} p{1.5cm} c >{\small}p{11cm}}
\toprule
\textbf{PID} & \textbf{Active Ingredient Inspiration} & \textbf{Species} & \textbf{Code} & \textbf{Participants' Think-aloud \& Related Behavior Descriptions} \\
\midrule
P1 & Shape allows air to flow over and around like a sail & Caryophyllales (seed) & {\textbf{S1}} & \pquote{Okay.  WindSail carrier. \textbf{Oh that's pretty cool!} Inspired by the aerodynamically shaped seeds of Caryophyllales, this bike rack utilizes a lightweight, sail-like structure that harnesses airflow to reduce drag...} \\

& &  & {\textbf{D1}} & \pquote{The sails are adjustable to snugly fit bikes, mimicking the efficiency of seed dispersal by wind... What does it mean by a sail} (\textbf{Asks \sys{} a question to explain how `sail' would work in the idea}) \pquote{That's a lot of information... Aerodynamic Shape. The sail-like structure of the WindSail Carrier is not just for aesthetic appeal; it serves a functional purpose by mimicking the shape of aerodynamically efficient seeds. \textbf{The shape allows air to flow over and around the bike rack... Oh okay now we're getting something.}} \\
\midrule
P4 & Appendages retract into an empty space & Turtle & \textbf{D2} & \pquote{Okay, so, the tortoise shell \textbf{made me think about} how things can be folded into empty space. That was the thing I got from the tortoise. There's empty space inside the shell and it can fold like it can take its feet into the shell. But it doesn't break the feet into simple pieces, or fold it like, roll it like, or anything like that. Just takes the feed into empty shell. So that's what I came up with, and then I started thinking about like, oh what does it mean to have a slot inside, and then I thought, airplane wheels, and the Alaska airline door incident, which made me think about the pin-release mechanism (that was supposed to hold the door).} \\
\midrule
P6 & Small, powerful thrusts for quick, upward propulsion and  directional changes & Lepidoptera (butterflies and moths) & \textbf{S1} & \pquote{Okay \textbf{that's pretty interesting}, the propulsion (mechanism) and changing directions... that could be relevant to changing directions on wide stairs.} \\
& N/A & N/A & \textbf{S2} & \pquote{okay so since I don't know these concepts from nature, \textbf{I need help in understanding whether} I can use that technology in this? So \textbf{it's hard to understand the relevance.}} \\
\midrule
P11 & Sliding and collapsing (like in a telescope) & {Armadillo} & \textbf{D2} & \pquote{Okay this shell that can collapse is an interesting mechanism. Like this \textbf{makes me think of a telescope}, like a \textbf{telescoping mechanism for sliding and collapsing}... so that could be a really interesting design space.} \\
\bottomrule
\end{tabular}
\caption{Transcripts of participants' think aloud and behavioral records in the \sys{} condition (`Participants' Think-aloud \& Related Behavior Descriptions') and how they were coded into different types of engagement patterns (`Code'). The bold-faced text in each row highlights the important signatures of the assigned code. Each row also contains the following: the participant ID (`PID'), the active ingredient description that participants found interesting / relevant (`Active Ingredient Inspiration'), and the associated species (`Species'). Exhibits in the baseline condition were similar, with the exception of tools participants interacted with.}
\label{table:ex_actions_and_codes}
\end{table*}

\begin{figure}[t]
    \vspace{-1em}
    \centering
    \begin{subfigure}[t]{.2\textwidth}
        \centering
        \includegraphics[height=4.2cm]{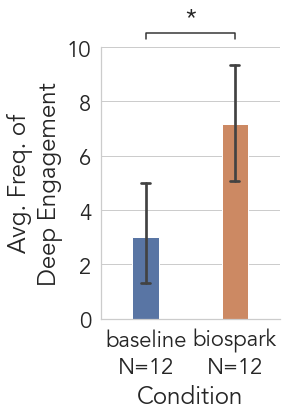}
    \end{subfigure}
    \quad
    \begin{subfigure}[t]{.2\textwidth}
        \centering
        \includegraphics[height=4.2cm]{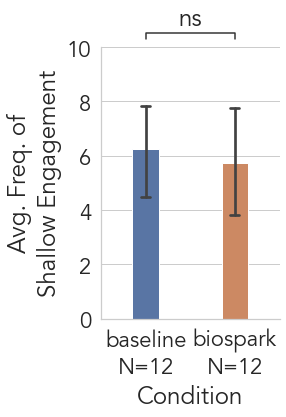}
    \end{subfigure}
    \vspace{-1.5em}
    \caption{
    (Left) The bar graph shows that the average number of deep engagement was significantly higher in \sys{} than in Baseline.; (Right) The bar graph shows that there were equally many shallow engagement types in both conditions.}
    \vspace{-2em}
    \label{fig:engagement_freq}
\end{figure}
To determine whether there was a difference between the frequencies of the two types of codes and conditions we conducted a $\chi^2$ test, finding a significant distributional difference (\chisq{12.93}{.0003}).
Follow up pairwise comparisons after Bonferroni correction showed a significantly higher frequency of deep engagement in \sys{} (M=7.2, SD=4.04) over the baseline condition (M=3.0, SD=3.30) (\tpaired{21.16}{-3.12}{=.01}, fig.~\ref{fig:engagement_freq}, left).
These results suggest that \sys{} promoted greater follow-up and idea generation than the baseline; we explore the usage of features that may have contributed to this difference further below.

No significant difference was found in the frequency of shallow engagement between conditions (Baseline: M=6.3, SD=3.28; \sys: M=5.8, SD=3.60, \tpaired{21.81}{.34}{=.74}, fig.~\ref{fig:engagement_freq}, right), suggesting that participants in both conditions encountered inspirations that they did not prioritize for follow up.
However, we found that positive utterances for shallow engagement (S1, \eg ``Interesting!'') were significantly higher compared to negative utterances (S2, \eg ``I don't see the relevance.'') in the \sys{} condition (M=3.1, SD=3.92) compared to the baseline condition (M=-.3, SD=4.16) (\tpaired{21.92}{-2.25}{=.046}).
Thus is it possible that even though they did not follow up on them, participants may have encountered even more inspirations of possible interest in \sys{}, making the findings of engagement above conservative estimates.
\vspace{-1em}
\subsubsection{Design constraint mentions}
\label{subsubsection:addressing_design_constraints}
We also explored how often participants mentioned design constraints mentioned in the brief, which are requirements to be considered for the idea to be useful (\eg being lightweight but durable, or versatile in accommodating different shapes).

Although we do not have direct measures for how well these constraints were addressed beyond the quality ratings discussed above, an indirect measure we explored was the number of times design constraints were mentioned in participants' ideas. We performed a two-tailed, independent samples t-test over ideas grouped by condition, finding that participants' ideas were significantly longer in the \sys{} condition (M=375.5, SD=96.15) over the baseline condition (M=141.9, SD=108.93, \tind{119.25}{-14.65}{<.0001}).
However, length alone may not represent how many different design constraints participants mention.
To examine constraint-specific content we extracted unique chunks from each idea description that corresponded to each design constraint using the \codett{gpt-4-turbo-preview} model, which showed satisfactory performance (details in Appendix~\ref{appendix:study_extract_constraints}).
From this we find that \sys{} users mention a significantly higher number of design constraints in their ideas (M=2.7, SD=1.01) than the baseline condition (M=1.6, SD=.78; \tind{163.27}{-8.45}{< .0001}), suggesting they may have considered design constraints more in their process.

\vspace{-1em}
\subsection{Challenges}
\xhdr{Controllability around what problem constraints are emphasized, and by how much}
In some occasions participants felt the AI's response was re-coursing to make connections to the original problem constraints, which seemed forced and adding less value \pquote{it's still trying to navigate the conversation towards the original topic and doesn't seem to be ``progressive'' enough to talk more in depth with specific areas} (P12). 
Similarly, P10 also said: \pquote{I think the AI is trying too hard here to make the connections. I'm more interested in the function of a climbing wheelchair, and want to focus on that first rather than the weight and foldability which are secondary challenges to me}.
However, sometimes this might potentially have a beneficial problem-reframing effect and serve as an entry into a new design space to explore: \pquote{I don't think these really addressed my questions exactly, but maybe [adapters] is what I'm looking for} (P5).

\xhdr{Supporting deeper idea development}
Participants also wished to have more support around developing ideas further with relevant technical specifications.
P10 said: \pquote{I'm interested in going deeper in the technical specs. I've read the brainstormed idea description, and now I want to build more confidence in it.} (P10).
P5 wanted a way to easily \pquote{extract interesting parts from an earlier information card and integrate them directly into the next sparks}, while also being able to \pquote{ask additional follow-up questions based on the previous Q\&A responses, without overcrowding the stream space} (P5).

Some participants wanted additional support for \pquote{comparing and contrasting} (P7) the ideas, to \pquote{easily highlight the strengths of each design and extract useful design features from it for integration into a new idea for mitigating anticipated challenges} (P5), beyond the current single-idea-focused trade-offs analysis.
Participants wished an easier way to organize the pros-and-cons table for multiple material candidates and comparison: \pquote{It would be great to be able to specifically consider different variations / combinations of different materials, titanium for the base and fiberglass for the body or cover} (P12), and have a way to perform an estimated quality analysis by swapping some features of an idea with those from other ideas or candidate materials: \pquote{How can I apply this to the cover portion of the bike-rack and what would happen if I replaced this with this other material?} (P6).

\xhdr{Idea organization and information space efficiency}
With many sparks and information cards generated, participants wanted a better mechanism for crowd and space management.
Participants engaged in grooming the space as desired (\eg \pquote{I'll eliminate some of the ideas so that I don't get overwhelmed} -- P10; \pquote{I don't have a lot of space over here, a bit frustrating to me} -- P11), and in making the ideas being actively pursued more easily accessible by prioritizing and ranking ideas: \pquote{Is there a way to bookmark some things (beyond the save-the-mechanism feature)? I think when you're researching multiple ideas, it would be useful to easily narrow down to a few interesting things for further consideration} (P5).
Participants also wished to have a way to perform this prioritization via \pquote{comparisons of this spark and this other spark in terms of strengths and weaknesses} (P7).

\xhdr{Other challenges}
Some participants noted the potential value of being able to switch between retrieval- vs. generation-based modes of getting the information cards.
For example, P11 commented that \pquote{(The baseline) is more intuitive for me because I can always switch between different tabs and ask questions to AI if I have to, or instead Google something, and not always invoke AI automatically} (P11).
Related to this, P10 commented that he wanted to \pquote{understand what the source is when new ideas are generated} (P10), suggesting further exploration is needed to understand how knowing which information sources contributed to the card content affects the perception of trust, which may become more important for further idea development and could be useful for related customization.

Moreover, participants commented on wanting to instruct the output structure as desired related to more or less details or conciseness, or more or less structure for skimming.
P10 said, \pquote{I also would like bullet points as opposed to long text or paragraphs for scannable and easier comparisons} (P10).
Similarly P11 commented that he would \pquote{like (the cards) to be as concise as possible... surfacing the design objectives over complete sentences, along specific criteria or ingredients -- these are the features and important stuff} (P11).

Together, these challenges point to limitations of the interaction paradigm proposed in \sys{}, and fruitful areas for future research around personalizing and improving the controllability in LLM-augmented analogical transfer.

\vspace{-1em}
\section{Discussion} \label{appendix:future_work}
We introduced \sys{}, a system exploring the idea of acting as a creativity partner in analogical innovation. \sys{} builds on insights from a design workshop and formative pilot study to support not only finding inspirations but also transferring inspirations into the target design domain and more deeply engaging with them during ideation. We found in a user study that the LLM-enabled features we explored in \sys{} -- generating and clustering inspirations, introducing sparks to help map the inspiration to the design problem, trade-offs to help users consider design constraints, and free-form chat to explore inspirations more deeply -- resulted in participants generating more ideas and exploring more different species without a significant decrease in diversity compared to a `gold standard' condition using AskNature inspirations and ChatGPT. Furthermore, \sys{} appeared to keep users in the flow of ideation, reduce the cognitive effort in transferring and adapting ideas, and help people engage more deeply in considering how they could use inspirations and the design spaces they unlocked.

One significant concern we had was that the features that were aimed at deeper engagement, such as sparks, might counter-productively decrease engagement and increase fixation because of how fleshed out the connections were in terms of articulating an entire, detailed design idea embodying the inspiration's mechanism in the target domain (\eg using spider silk for lifting a wheelchair or creating a ramp). The higher the fidelity of an inspiration the more it may incur fixation and direct use rather than creative adaptation~\cite{vasconcelos2016inspiration}.
In our evaluation we cannot definitively reject this possibility.
While our analysis of interaction logs and think-aloud transcripts suggests that participants actively engaged with the mechanism inspirations -- recognizing their relevance, deepening their understanding, and transferring them to the target design domain -- further investigation is required to unpack the effects of AI-enabled content generation features, such as the spark, trade-off, and Q\&A cards, on the idea generation process.

With this caveat, we highlight some factors that may warrent future research. First, although the AI appeared to reduce the cognitive load of mapping the idea to the design space, the decision to do the mapping in the first place was user-driven, prompted by them saving an inspiration.
Thus before seeing the AI mapping they needed to notice something interesting or relevant about the inspiration, even if they didn't fully make the connection between it and the problem domain. This self-driven curiosity and agency may influence how users engage with sparks and trade-off cards. Future systems could explore which user actions and levels of agency are necessary to foster a sense of ownership and spur initial engagement with inspirations.

Another factor that might have driven engagement was perceived ownership of inspirations. Previous work has identified that ownership and attribution are key elements of human-AI collaboration~\cite{peng2020exploring}. In our study we noticed participants making attribution statements about the sparks, such as ``\pquote{That's not really my idea, [it's] ChatGPT's idea but okay}''.
A common theme among participants was discussing how they modified the sparks to make ideas more their own and to avoid ``plagiarism'', even though they were told they could use the sparks as they wish.
It's unclear why designers in other studies finding fixation when using LLM and generative AI did not similarly adapt and riff on ideas in order to build ownership, but a possibility that might be explored is that an integrated system that frames AI-generated ideas as intermediate products, as we do in \sys, might be more effective at promoting deeper engagement than an unstructured system or one where they claim prominence as more final products.

\xhdr{Fixation vs. Contextualization} Another concern we had in our system design was whether contextualizing the system interface features in the design problem would lead to narrowing of the design ideas users would explore.
Sparks, trade-off cards, and the free-form chat interface were all contextualized with the source design problem, with the goal of reducing the cognitive effort needed for users to engage with the details of the inspirations relevant to their goals.

This largely appeared to hold true, with users finding the contextualization useful, and even sometimes ``magical''. While they were technically able to do this with the baseline system and sometimes did (\pquote{It feels like I got the seed like the very starting point idea from AskNature and then generating actual ideas from it was done by ChatGPT, like translating the seed into actual ideas.}), the efficiency of the built-in contextualization was frequently mentioned as useful as a jumping off point rather than replacing cognitive work (\eg~\pquote{it was able to produce things without me having to like prompt it. And I think that allowed me to spend more time, maybe thinking about specific connections between the mechanisms and the design features it was suggesting.}). Participants also commented on the reduction of cognitive effort in transferring inspirations, as P1 described: \pquote{I just felt like it was able to produce things without me having to like prompt it. And I think that allowed me to spend more time, maybe thinking about specific connections between the mechanisms and the design features it was suggesting. Whereas... with [ChatGPT] I felt like I had to spend more time like doing to allow it to actually help, but with [\sys] I felt like the AI system already knew what I needed, so it saved that step.}. P5 felt it was \pquote{easy to highlight the strengths of each design and extract useful design features from it for integration into a new idea for mitigating anticipated challenges}, and that \pquote{(The sparks are) very specific... how did it know? How are these so specific to bike rack design? Wait, it's already pitching me different ideas. That's so cool} (P5).

In contrast, the baseline system seemed to result in significant cognitive effort, both physically not having the context of the design problem, and attentionally in not keeping them focused.
For example, P12 commented that \pquote{This [AskNature.org] article could be tailored more to the bike rack design, instead it's about stuff that's in motion... so it's hard to know what the relevance is} (P12).
Without the persistent context, and source inspirations on AskNature disconnected from the target design problem, participants felt the inspirations in the baseline \pquote{didn't tell me about too much in terms of what insights I can draw to support (the scenario described by the design brief)} (P1) and that they had to engage in multiple back-and-forth's with ChatGPT to glean transferrable insights and ideating specific design ideas based on them, which was time-consuming without the persistent context and consistent user guidance: \pquote{(Types in ChatGPT) it's giving me specific animals (instead of transferrable mechanisms) so I'm going to pause the generation. Let me see... (Types in ChatGPT) `what shape?'. Oops that didn't work, let me try again. `What shape that could be useful for industrial design?'} (P12). For P4, AskNature.org felt \pquote{like a generic platform} and that building off of its material using AI felt like \pquote{asking just random questions to ChatGPT, while in \sys{} I was asking related questions} (P4).

Overall, our results suggest a more nuanced consideration of context and fixation than previously considered, in which helping users reduce cognitive load throughout the analogical innovation process while positioning AI suggestions as intermediate products in the system flow, could represent a profitable paradigm to explore.

\xhdr{Limitations}
Our findings and analysis have several limitations. First, our access to professional designers working in large organizations was limited to a design workshop. Our formative study and user studies involved heterogeneous participant pools, including freelance designers recruited from UpWork and design and PhD students recruited from an arts college. These participants may not fully represent the broader population of design professionals.

Second, alternative interpretations of our data are possible based on the operationalization of the measures we used. Creativity and ideation measurement spans multiple disciplines, and our chosen metrics -- focused on engagement with inspirations, as well as the quality, quantity, and diversity of ideas -- reflect a specific subset of this research. Notably, our approach to measuring idea diversity, such as extracting species' names from ideas or relying on a particular LLM in our computational pipeline, may introduce biases.

Additionally, our study protocol, while informed by professional design practices, involved artificial scenarios and time constraints that may not mirror real-world design contexts. The self-driven curiosity and agency observed in participants' engagement with AI-generated inspirations may differ in professional settings, where constraints and priorities vary. Future studies embedding such systems into designers' ongoing work or making them publicly available for broader use could yield valuable insights into practical benefits and remaining challenges.

Our evaluation suggests that while the system reduces cognitive load during the analogical ideation process, the decision to map inspirations to the design space remains user-driven, requiring participants to identify relevant or interesting aspects of inspirations. The interplay between AI support and user agency as well as their effects on design fixation warrants further exploration to better understand the balance between automation and human creativity in fostering meaningful and innovative design processes.

Finally, our work aims to support only the ideation portion of the design process, while a significant part of the design process also involves prototyping and evaluation. Further work is needed to extend these results to incorporating later stages of the design process.

\section{Conclusion}
In this work we present \sys, a creativity partner for analogical design.
\sys{} proposes a new interaction paradigm for reducing the cognitive effort in finding, recognizing, mapping, and creatively adapting diverse inspirations from distant domains.
Future work remains in this area to generalize our diversification structure and goal-driven inspiration discovery pipeline for generating analogical inspirations from different domains, and in further personalizing, controlling, and customizing support for user recognition and idea development beyond what was explored here to accelerate and materialize high-impact innovations.
We imagine a future in which engineers and designers could find inspirations based on deep analogical similarity that move beyond domain boundaries to drive innovation across fields.

\begin{acks}
This research was supported by the Toyota Research Institute and the Office of Naval Research.
\end{acks}

\bibliographystyle{ACM-Reference-Format}
\bibliography{main}

\appendix

\section{Inspiration Design Probe Study Prototype Implementation}
\subsection{Representative Mechanism Image Curation} \label{appendix:design_inspiration_probe_image_search}
To aid designers' visual understanding of and pique curiosity for biological-analogical mechanisms, we retrieve representative images for corresponding textual mechanism descriptions.
We use Google Custom Search\footnote{\url{https://developers.google.com/custom-search/v1/overview}} with queries as ``\code{[organism name]:[mechanism description]}'' and the file type set to images and the safe search mode enabled.
We choose the first result of Custom Search as the visual representation of each mechanism.

\subsection{Interface} \label{appendix:design_inspiration_probe_interface}
\begin{figure*}[th]
    \vspace{-1em}
    \includegraphics[width=\linewidth]{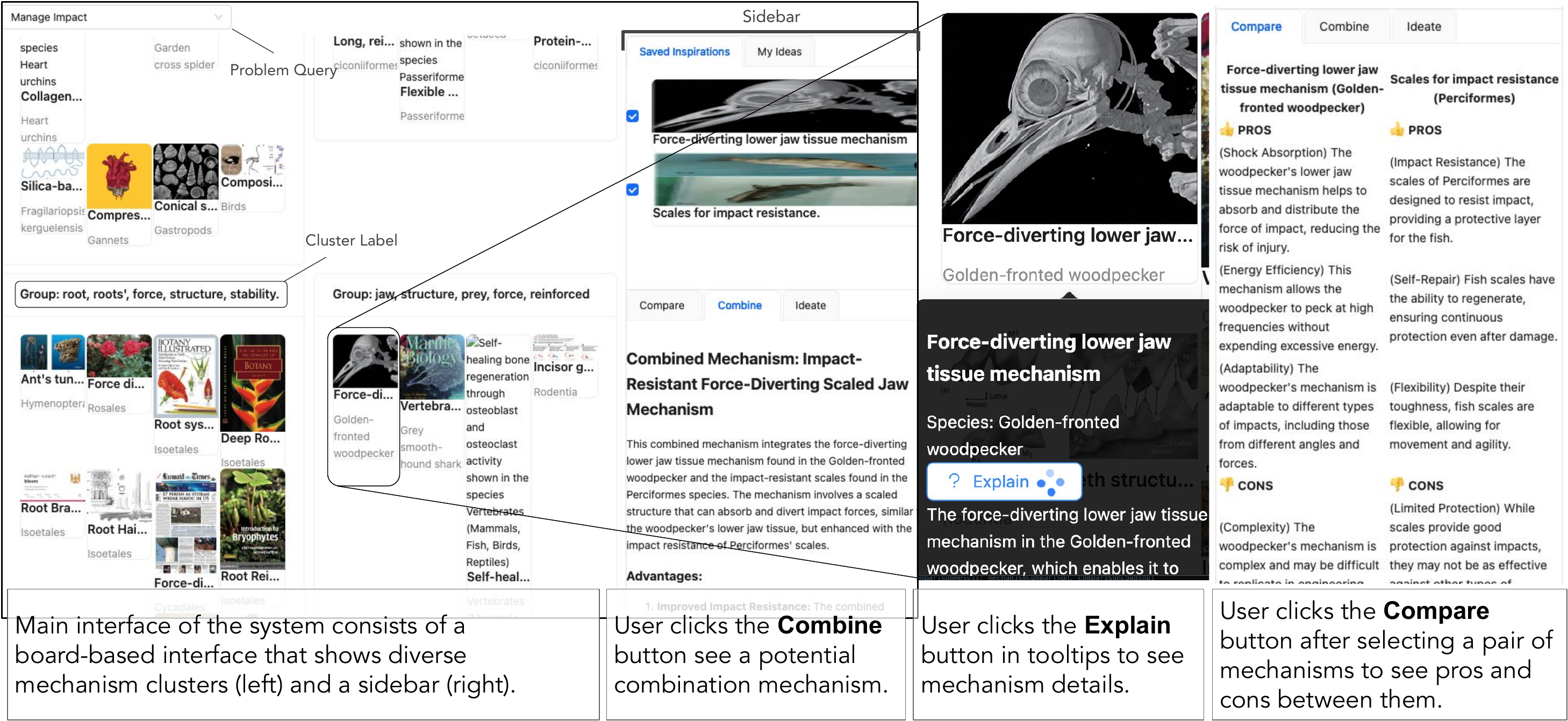}
    \vspace{-1.5em}
    \caption{Inspiration design probe interface and a subset of available interaction features. The interface consists of a left pane that shows clusters of semantically similar mechanisms that was scrollable, and a right pane that included a holding tank for user-saved mechanisms. When the user checks two of the saved mechanisms in the holding tank and clicks one of the tabs underneath, the system generated the corresponding content, such as the comparison of two mechanisms in a pros-and-cons table, a new idea that combines the two mechanisms, and the `Ideate' button that provided critique on the `Combine' idea.}
    \label{fig:design_inspiration_probe_interface}
\end{figure*}
To facilitate designers' understanding and synthesis of mechanism inspirations, we develop several interaction features available on the probe interface (fig.~\ref{fig:design_inspiration_probe_interface}).
The \textbf{Explain} button is located in tooltips that pop up when the user places the mouse over on a mechanism card in the board UI (fig.~\ref{fig:design_inspiration_probe_interface}, first panel).
When the user clicks on the button, \sys{} sends a prompt to GPT4 requesting elaboration of the interacted mechanism and the organism in the context of the chosen engineering design problem.
The \textbf{Compare} tab is located in the control bar of the sidebar of the interface.
To use this, users need to first click on (at least) two mechanism cards from the left, saving them to the `saved inspirations' panel at the top of the sidebar.
There, users can check any two of the saved mechanisms they wish to compare.
\sys{} sends a prompt to GPT4 when the user clicks on the tab, requesting comparison of pros and cons between the two mechanisms in the context of the chosen engineering problem.
The result is formatted into a markdown table, with each mechanism as the header followed by pros and cons rows detailing each point.
The \textbf{Combine} tab is also located in the control bar of the sidebar in the interface.
Similarly with Compare, users can check two saved mechanisms they wish to see combined.
\sys{} sends a prompt to GPT4 then requesting elaboration of a mechanism that combines the two selected mechanisms, and explain its potential advantages in the context of the chosen engineering problem.
The result is also formatted into a markdown page using section title and headers for demarcating the content.

\section{\sys{} Implementation Details}
\subsection{Visual Representation of Inspirations}
To support the design goal of visual representations of inspirations we were inspired by AskNature.org, a popular web repository for bio-inspired design.
AskNature pages are designed to include a prominent close-up and centered portrait of a species that creates a striking visual and invokes curiosity. 

We retrieve animal images using Google Search and Adobe Stock Images APIs, and use GPT-4V to rank the quality in terms of the visual focus and the potential value for mechanism understanding.
The first author developed a prompt (fig.~\ref{fig:ranking_species_images_prompt}) by reviewing the ranked images, their scores, and rationale for 10 species.
Applying this led to a consistent pattern of reasonable visual processing and instruction-following, as demonstrated in fig.~\ref{appendix:sample_scores_and_justifications} to be useful.

\subsubsection{Prompt Design}
\label{appendix:ranking_species_images}
The developed prompt is shown in fig.~\ref{fig:ranking_species_images_prompt}.
\begin{figure*}[htbp]
\noindent\rule{\textwidth}{0.4pt}
\begin{lstlisting}
[User Message]
Judge each image based on how clearly it shows the real species (i.e., photos focusing on one instance of the species in the wild is better than cartoons, drawings, or species photographed in the distance)
{species} and contains visual details that help viewers understand the following biological mechanisms:
===
{formatted_mechanisms}
===
For each image given, reply with a number between 0 and 100 as its "score", where a higher number represents a higher quality of the picture,
and also provide rationale for your decision in "rationale".
Output a list, with the following format. Exclude any other character than the comma between dictionaries in the list:
[{{"score": "50", "rationale": "..."}}, ...]
\end{lstlisting}
\noindent\rule{\textwidth}{0.4pt}
\vspace{-1.5em}
\caption{The prompt used to score each species image with respect to its visual focus on the species and how helpful it might be for understanding the specific mechanism of the species.}
\label{fig:ranking_species_images_prompt}
\end{figure*}

\subsubsection{Sample Scores and Justifications} \label{appendix:sample_scores_and_justifications}

\begin{figure}[t]
    \centering
    \begin{subfigure}[t]{.12\linewidth}
        \includegraphics[width=\linewidth,height=\linewidth,clip,trim=0cm 0cm 0cm 0cm]{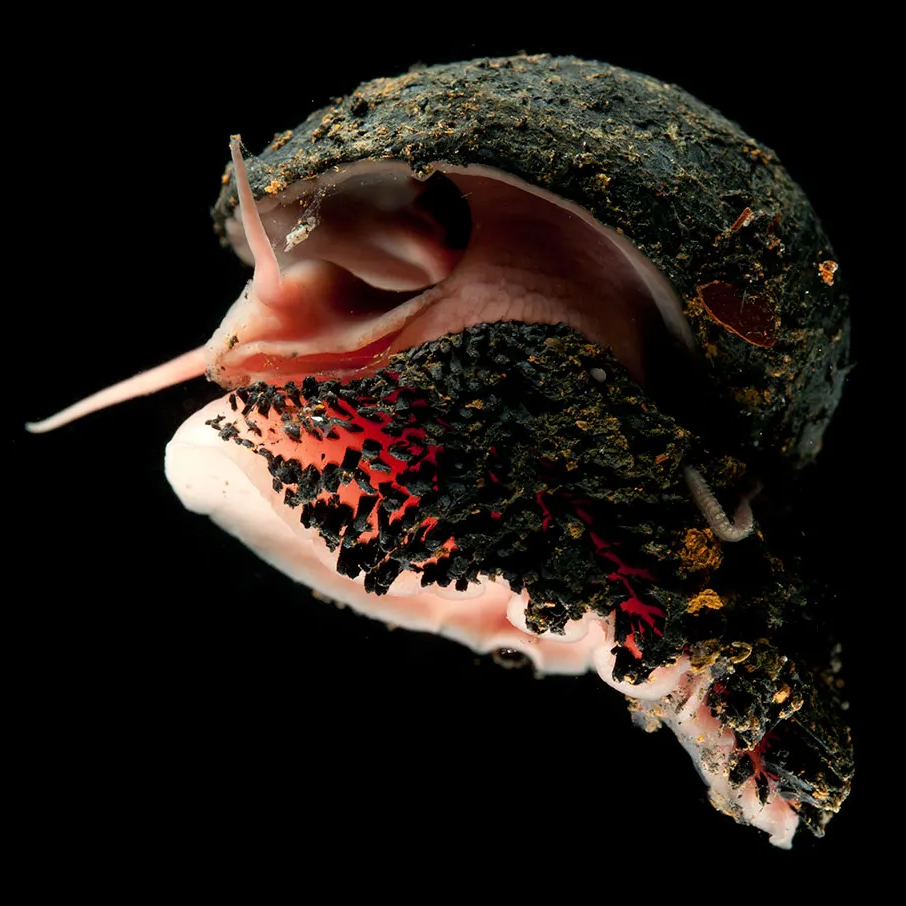}
    \end{subfigure}
    \begin{subfigure}[t]{.12\linewidth}
        \includegraphics[width=\linewidth,height=\linewidth,clip,trim=0cm 0cm 0cm 0cm]{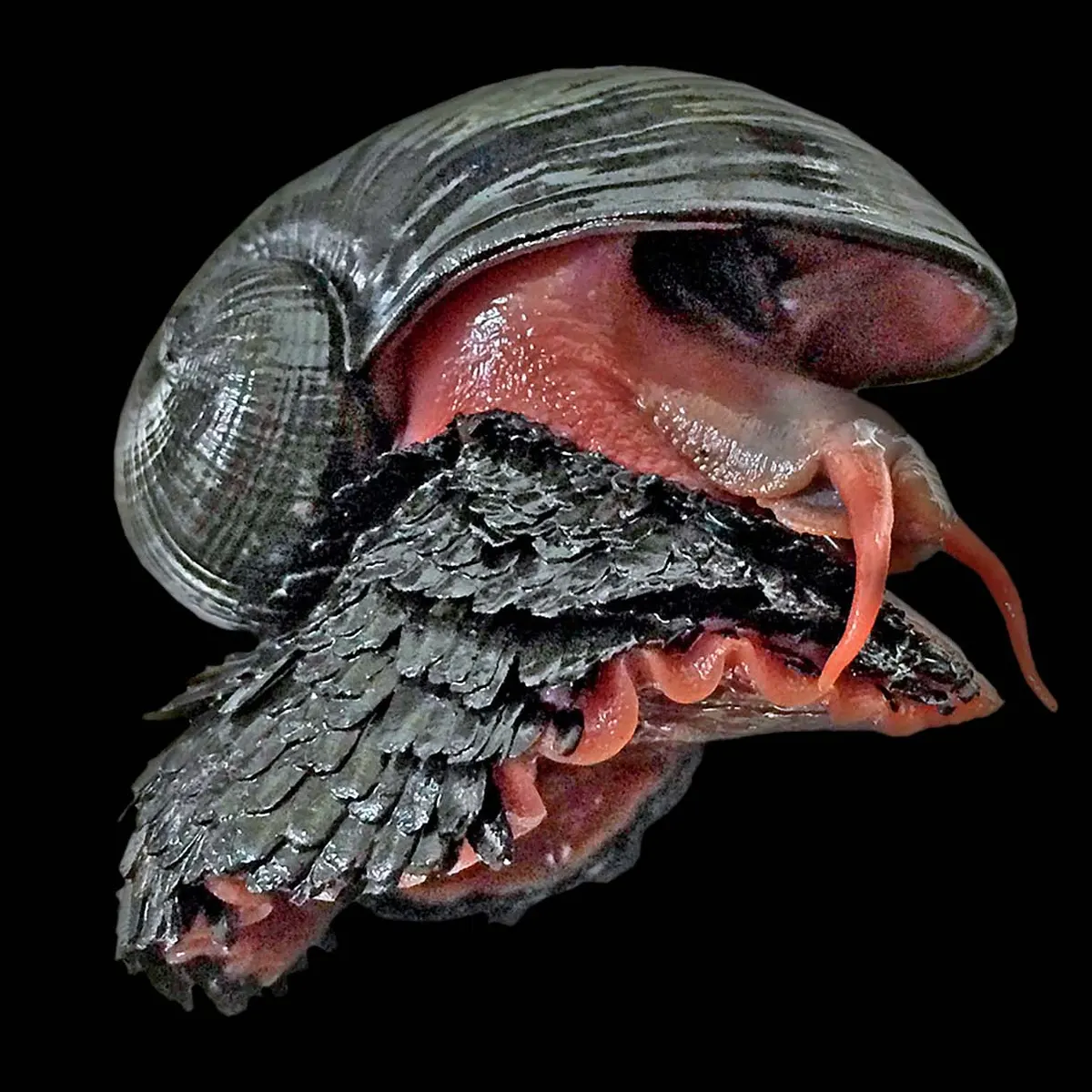}
    \end{subfigure}
    \begin{subfigure}[t]{.12\linewidth}
        \includegraphics[width=\linewidth,height=\linewidth,clip,trim=0cm 0cm 0cm 0cm]{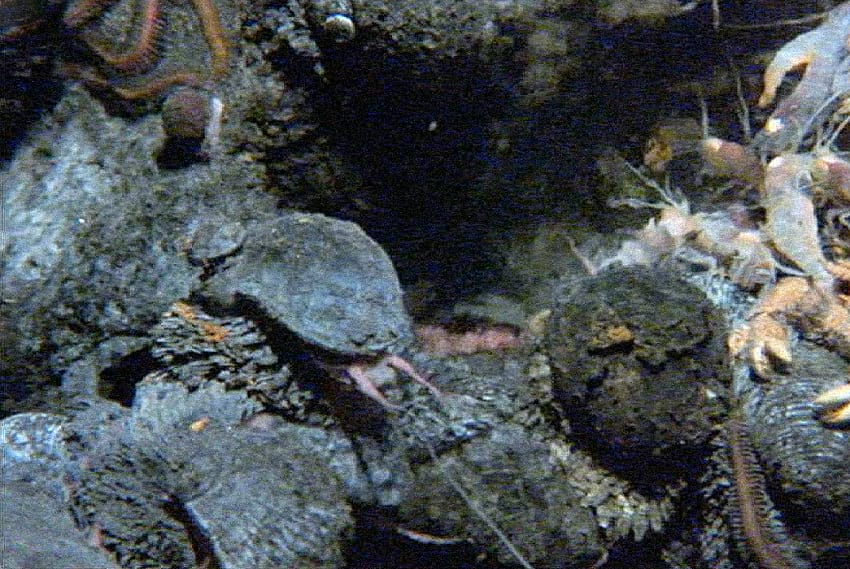}
    \end{subfigure}
    \begin{subfigure}[t]{.12\linewidth}
        \includegraphics[width=\linewidth,height=\linewidth,clip,trim=0cm 0cm 0cm 0cm]{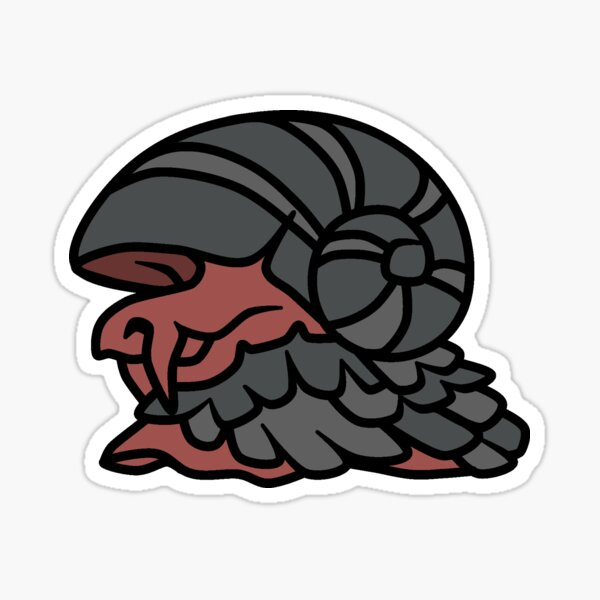}
    \end{subfigure}
    \vspace{-1em}
    \caption{The top and bottom two images ranked by GPT-4V (scores from the left: 95, 92, 30, 10) for `Scaly-foot snail'.}
    \label{fig:ranking_species_images_ex}
    \vspace{-1em}
\end{figure}
For the example shown in fig.~\ref{fig:ranking_species_images_ex}, the model scored and ranked the images as top-2 and bottom-2 candidates (from left-to-right) for the species `Scaly-foot snail' as follows:
\begin{lstlisting}[style=prompt]
Top-1 | Score: 95 | Rationale: This image provides an excellent close-up view of the scaly-foot snail, highlighting intricate details of the iron-infused sclerites and the shell's texture which are indicative of its biological mechanisms.

Top-2 | Score: 92 | Rationale: This image offers a clear and detailed close-up of the scaly-foot snail, clearly showcasing the iron-infused sclerites and the unique appearance of its iron-mineralized shell, which are valuable for understanding its biological mechanisms.

Bottom-2 | Score: 30 | Rationale: Although this underwater image shows the scaly-foot snail in its natural environment, the details necessary to comprehend the biological mechanisms such as the sclerites and shell are obscured and not clearly visible.

Bottom-1 | Score: 10 | Rationale: The image is a stylized, cartoon-like representation of a scaly-foot snail, lacking detailed visual information about the species' biological mechanisms such as the iron-infused sclerites and iron-mineralized shell.
\end{lstlisting}

\subsection{Goal-driven Inspiration Discovery Pipeline Implementation}
\subsubsection{Structuring AskNature blog posts into seed  problem-mechanism-organism schemas} \label{appendix:schemas_from_asknature}
To source a set of diverse, high-quality biological mechanisms for a given problem, \sys{} starts from a seed set of expert-curated biological mechanisms on AskNature (fig.~\ref{fig:backend_tree_generation}, Step 1). 
AskNature.org provides a curated list of organisms with detailed descriptions of their unique strategies to functional problems (\eg `Manage Impact', `Modify Speed').
The organisms and strategies can be grouped by function and viewed as a list.
To curate a seed set of high-quality mechanisms, we first choose a functional problem $p$ predicted to be highly relevant to automobile designers, excluding irrelevant functions such as `Adapt Behaviors', `Adapt Genotype', `Coevolve', `Maintain Community'
We access the sub-list of organisms $o \in O$ and strategies posted to $p$ on AskNature's group-by-function page by parsing the \code{HTML} code using the \code{BeautifulSoup} package on \code{Python}. We then access the blog post for each organism-strategy page using the parsed URL and parse the returned \code{HTML} page to get the title, description, and references (if available).

At this stage, the returned unstructured text is yet to contain a succinct mechanism description.
Furthermore, we found that some blog posts do not contain any body text despite having a title and are accessible via the URL.
Some of these missing blog posts indicated that they are in maintenance and/or planned to be updated.
To structure the raw blog post text $\text{AskNature}_{(o, p)}$, we prompt GPT4~\cite{openai2023gpt4} to succinctly describe (\ie using 12 words or less) the core mechanism (\ie excluding the qualities or effects, and focusing on mechanisms with engineering design implications), given $(o, p)$ (if blog post text is missing) or $(o, p, \text{AskNature}_{(o, p)})$.
The returned mechanism description $m$ along with the function description makes up the problem-mechanism schema for each organism: $\{o \in O|(p, m, o)\}$.

\subsubsection{Iteratively expanding mechanisms dataset by traversing constructed taxonomic trees} \label{appendix:iterative_structured_expansion}
Using each schema as a seed, we iteratively prompt GPT4 to find relevant mechanisms for the given mechanism and problem, using an even mixture of breadth- and depth-focused expansion strategies (fig.~\ref{fig:backend_tree_generation}, Step 2).
To enable structured diversification of organisms and their mechanisms beyond prior work that relied on token-level manipulation or na\"{i}vely prompting LLMs, we guide LLMs {how} and {where} to expand by leveraging organism taxonomic hierarchies.
At each iteration of expansion (fig.~\ref{fig:backend_tree_generation}, Step 2), we aggregate the organisms represented in found mechanisms up to that point, and construct a taxonomic tree featuring seven levels of hierarchy on Tree of Life: \{\code{domain, kingdom, phylum, class, order, family, genus, species}\}, where \code{domain} representing the highest level and \code{species} representing the lowest level on the hierarchy.

Given this tree, we aim to identify sparsely populated branches for expansion.
We cut the tree at a given reference expansion level (\eg \code{class}), and sort the taxonomic ranks (nodes) on that level by the number of its immediate children nodes\footnote{Alternatively, the entire size of the subtree, rather than immediate children, could be used for sorting}, in an increasing order. 
For performance, we batch 10 prompts to send to GPT4 for expansion.
For half of the prompts, we instruct \textbf{breadth-first expansion} which asks GPT4 to first identify \textit{sibling} nodes at the given reference taxon level and existing nodes (up to 50 most populated nodes).

For example, the prompt asks ``come up with a few biological \code{classes} not in \{\code{...names of excluded classes...}\}''.
The breadth-first expansion prompt then instructs GPT4 to repeat the following: 1) Choose one taxon from the list it came up with; 2) Succinctly describe (\ie using 14 words or less) new mechanisms $m$ related to a problem $p$.
For the rest of the prompts, we instruct \textbf{depth-first expansion} which asks GPT4 to first identify a new \textit{children} node at the given reference taxon level and existing children nodes (up to 50 randomly sampled children).
For example, the prompt asks ``come up with a few biological \code{families} in \code{order} \code{araneae} that are not any of \{\code{araneidae}, ...\}''.
The depth-first expansion prompt then instructs GPT4 to repeat a similar procedure as breath-first expansion.
The prompt details are provided in fig.~\ref{fig:depth_expansion_prompt} (the depth-focused expansion prompt) and fig.~\ref{fig:breadth_expansion_prompt}.
In the prototype system, we run 10 batches for expansion to construct dataset of mechanisms for each problem.

The returned list of mechanisms and organisms text are then fed into the second GPT4 prompt for structuring them into a list of \{\code{mechanism, organism}\} dictionaries.
Finally, using each organism name, we prompt GPT3.5-turbo to retrieve the seven-level taxonomic hierarchy, based on our model evaluation result showing its high accuracy (Appendix~\ref{appendix:accuracy_taxonomy_construction}).

\begin{figure*}[htbp]
\noindent\rule{\textwidth}{0.4pt}
\begin{lstlisting}
[System Message]
You are an expert biologist who knows species and their taxonomic hierarchy in detail.
You can also come up with diverse problem-solving strategies found in nature relevant to engineering design problems.
Do the following step-by-step.
\end{lstlisting}
\noindent\rule{\textwidth}{0.4pt}
\begin{lstlisting}
[User Message]
1. Come up with a few biological {lower-taxon-plural} **IN** the {taxon} "{term}" AND **NOT** {exclude-user-prompt}
2. Select one {lower-taxon_singular} from the list you came up with.
3. Come up with short descriptions (up to 14 words or less) of new mechanisms found in the selected {lower-taxon-singular} that are applicable to the challenge of "{prob}".
4. Repeat step 2 and 3 for each selected {lower-taxon-singular} and think step-by-step. Number each step in your thinking and make it as short as possible.
\end{lstlisting}
\noindent\rule{\textwidth}{0.4pt}
\vspace{-1.5em}
\caption{The prompt used for depth-focused expansion of the mechanism dataset. The ``lower-taxon-singular'' or ``lower-taxon-plural'' is the singular and plural name of the subsequent level on the tree-of-life hierarchy, of the level ``taxon'', respectively. The ``term'' is the name of the selected taxon. The ``exclude-user-prompt'' includes previously generated ``taxon'' names which are used to instruct the LLM to avoid duplicate generation. The ``prob'' and ``src-mech'' contain the problem and mechanism schemas to constrain generation.}
\label{fig:depth_expansion_prompt}
\end{figure*}

\begin{figure*}[htbp]
\noindent\rule{\textwidth}{0.4pt}
\begin{lstlisting}
[System Message]
{same as in the depth-focused expansion prompt}}
\end{lstlisting}
\noindent\rule{\textwidth}{0.4pt}
\begin{lstlisting}
[User Message]
1. Come up with a few biological {taxon-plural} **NOT IN** the excluded {taxon-plural} below:
{exclude-user-prompt}
2. {same as in the depth-focused expansion prompt}
3. {same as in the depth-focused expansion prompt}
4. {same as in the depth-focused expansion prompt}
\end{lstlisting}
\noindent\rule{\textwidth}{0.4pt}
\vspace{-1.5em}
\caption{The prompt used for breadth-focused expansion of the mechanism dataset. See the depth-focused expansion prompt (fig~\ref{fig:depth_expansion_prompt}) for parameters descriptions.}
\label{fig:breadth_expansion_prompt}
\end{figure*}

\subsubsection{LLM-based Taxonomy Construction \& Accuracy Evaluation} \label{appendix:accuracy_taxonomy_construction}
The main process in our diversification strategy is iterative construction of taxonomic trees at each stage of expansion with a set of problem-mechanism schemas and corresponding organisms $\{o \in O|(p, m, o)\}$ curated (in case of AskNature seeds) or generated up to that point.
To construct the trees, the taxonomic hierarchy of each organism needs to be known.
Here, we restrict our tree construction to seven levels of depth, ranging from the highest to lowest levels: \code{domain, kingdom, phylum, class, order, family, genus, species}. 
These levels provide considerable branch-switching opportunities for diversification, through significant changes in the number of members between levels and within each level of the hierarchy.
For example, while the highest level \code{domain} consists of three members, Bacteria, Archaea, and Eukarya, there are estimated 8.7M species in the world~\cite{sweetlove2011}. 
The next level on the hierarchy, \code{Genus}, has an estimated number of 310K members~\cite{rees2020all}, while the number in the subsequent level, \code{families}, is estimated at 8K~\cite{mora2011many} in 2011.
The number of known species for each node on the hierarchy also changes considerably, further contributing to the diversification opportunities.
For example while most non-avian reptile genera have only 1 species each, insect genera such as \textit{Lasioglossum} and \textit{Andrena} have over 1,000 species each, while the flowering plant genus, \textit{Astragalus}, contains over 3,000 known species~\cite{wiki:genus}.

Our initial exploration of suitable approaches to retrieve organism taxonomies involved using available resources such as the Global Biodiversity Information Facility API\footnote{\url{https://www.gbif.org/developer/species}}, Catalogue of Life~\cite{banki2023}, or the Encyclopedia of Life~\cite{eol}, where canonical species names were retrieved from the Darwin Core List of Terms\footnote{\url{https://dwc.tdwg.org/list/\#dwc\_Organism}} for corresponding organisms in problem-mechanism schemas.
However, the limited coverage, data consistency, and API availability of these tools prevented their adoption.
On the other hand, Wikipedia provides scientific classification for some of the organism articles (for example in the Pomelo article\footnote{\url{https://en.wikipedia.org/wiki/Pomelo}}, taxonomic names for \code{Kingdom, Clade, Order, Family, Genus,} and \code{Species} are available in the `biota' information box that appears on the right-hand side of the page).
However, this data was not readily available for scalable generation.

\xhdr{Procedure}
LLMs may provide an alternative solution to the limitations of existing approaches for retrieving the taxonomic hierarchy for a given organism name.
To test this idea, we curated 90 gold taxonomies using Wikipedia that have complete information in the `biota' scientific classification info box (the complete list of 90 organism names can be found in Appendix~\ref{appendix:complete_list_of_organisms}).
For each organism, we prompted LLMs with each organism name zero-shot using the chat completions API endpoint\footnote{\url{https://api.openai.com/v1/chat/completions}} using each model key.
The prompt used for taxonomy generation for LLMs can be found in fig.~\ref{fig:taxonomy_prompt}.
\begin{figure*}[htbp]
\noindent\rule{\textwidth}{0.4pt}
\begin{lstlisting}
[System Message]
You are an expert biologist who knows species and their taxonomic hierarchy very well. Follow the instructions to the letter.
- Return the scientific term for each taxonomic rank the species belongs to.
- Enclose keys and values using double quotes ("...") and format them into a Python dictionary.
- Use the taxonomic ranks as keys and corresponding scientific terms as their values.
- Do not add any other text.
\end{lstlisting}
\noindent\rule{\textwidth}{0.4pt}
\begin{lstlisting}
[User Message]
What {"domain", "kingdom", "phylum", "class", "order", "family", "genus"} does "{organism}" belong to? Format your reply into a Python dictionary.
\end{lstlisting}
\noindent\rule{\textwidth}{0.4pt}
\vspace{-1.5em}
\caption{The prompt used to generate the taxonomy of each organism.}
\label{fig:taxonomy_prompt}
\end{figure*}
Once the hierarchy data is generated, we lower-cased the rank names for consistency.
\begin{table*}[t!]
    \centering
    \begin{tabular}{p{1.8cm} p{1.5cm} p{1.5cm} p{1.5cm} p{1.5cm} p{1.5cm} p{1.5cm} p{1.5cm}}
    \toprule
    \textbf{Model} & \textsc{Domain} & \textsc{Kingdom} & \textsc{Phylum} & \textsc{Class} & \textsc{Order} & \textsc{Family} & \textsc{Genus} \\
    \midrule
    \multirow{2}{*}{\code{GPT4}} & 100\% (90/90) & 100\% (90/90) & 100\% (90/90) & 100\% (90/90) & 96.7\% (87/90) & 94.4\% (85/90) & 98.9\% (89/90) \\
    \multirow{2}{*}{\code{GPT3.5-turbo}} & 100\% (90/90) & 100\% (90/90) & 100\% (90/90) & 100\% (90/90) & 95.6\% (86/90) & 95.6\% (86/90) & 93.3\% (84/90) \\
    \bottomrule
    \end{tabular}
    \caption{The accuracy of zero-shot taxonomy generation using only the organism name.}
    \vspace{-2em}
    \label{table:taxonomy_generation_accuracy}
\end{table*}

\xhdr{GPT4's Accuracy}
We find that GPT4's zero-shot taxonomy generation accuracy to range between 94.4\% and 100\% (Table~\ref{table:taxonomy_generation_accuracy}).
The lowest accuracy was observed in the \code{family} taxonomy, followed by \code{order} (96.7\%) and \code{genus} (98.9\%).

\xhdr{Error analysis} \label{appendix:taxonomy_generation_error_analysis}
We find that some error cases in taxonomy generation could be attributed to recent changes in classification in the literature.
For example, both GPT4 and GPT3.5-turbo models classified naked mole-rats as then literature-accepted `Bathyergidae' for their family, same as other African mole-rats.
However, more recently naked mole-rats were placed in a separate family, Heterocephalidae~\cite{Naked_mole_rat}.

Among the error cases overlapping between the two models, we found cases that either the GPT3.5-turbo or the GPT4 model wins over the other (\eg for `hummingbird', GPT3.5-turbo generated `archilochus' as its genus whereas GPT4 generated `various'; for `boxer crab', GPT3.5-turbo generated `hymenoptera' which is an order of insects, whereas GPT4 generated `decapoda', which is the correct order).
In other cases, both models outputted similarly incorrect answers, for example for `sea snail', GPT3.5-turbo generated `neogastropoda' whereas GPT4 generated `archaeogastropoda' (the Wikipedia gold answer was `lepetellida').

\xhdr{System Optimization: GPT3.5-turbo's Accuracy}
We find that GPT3.5-turbo has comparable accuracy levels with GPT4 in zero-shot taxonomy generation. 
The highest misaglignment occurred in \code{genus}, with a 6.67\% error rate (equivalent to 6 out of 90).
Appendix~\ref{appendix:taxonomy_generation_error_analysis} provides a further qualitative error analysis of models' comparative performance.
Based on these results, we opted for the more efficient GPT3.5-turbo model in our pipeline.
We leave further exploration of the capabilities of smaller, fine-tuned base LLMs, with implications for LLM cascade\footnote{LLM cascade refers to a system design approach that adaptively chooses optimal LLM APIs for a given query. Smaller, task-specific LLMs are regarded as optimal when they exhibit higher or similar levels of performance compared to models that are orders of magnitude larger~\cite{dohan2022language}, with all else being equal.}, to future work.

\xhdr{Complete List of Organisms Used for Taxonomy Generation} \label{appendix:complete_list_of_organisms}
\texttt{\{`spider monkey', `prairie dog', `garden tiger moth', `african sacred ibis', `argiope argentata', `ostrich', `groundhog', `danio rerio', `gannet', `deer', `cattle', `glyptodon', `alligator snapping turtle', `leopard', `arctic ground squirrel', `cormorants and shags', `bears', `squirrels', `herons', `european badger', `golden silk orb-weaver', `aardvark', `seahorses', `banner-tailed kangaroo rat', `hyenas', `pink fairy armadillo', `giant otter', `bighorn sheep', `hippopotamus', `california ground squirrel', `european bee-eater', `beech marten', `leopard gecko', `tailorbird', `testudinidae', `emperor penguin', `northern pike', `giant clam', `stoat', `horse', `nutria', `tree-kangaroo', `giraffe', `guinea baboon', `ferret', `bonytail chub', `baya weaver', `brook trout', `pelican', `mallard', `roseate spoonbill', `mountain weasel', `pocket gophers', `lybia edmondsoni', `giant anteater', `common raccoon dog', `dewdrop spiders', `armadillo girdled lizard', `arctic fox', `bison', `swordfish', `bald eagle', `chimpanzee', `asbolus verrucosus', `sperm whale', `abalone', `golden jackal', `hornet', `zebra', `orangutans', `peregrine falcon', `atlantic cod', `burrowing owl', `african wild dog', `maned wolf', `honey bee', `naked mole-rat', `echidnas', `bowerbirds', `rhinoceros', `beaver', `bombyx mori', `common box turtle', `hummingbird', `domestic sheep', `wolverine', `raccoon', `evergreen bagworm', `pig', `muskrat'\}}

\subsection{Active Ingredient Extraction} \label{appendix:active_ingredient_extraction}
To extract active ingredients from the outputs of earlier diversification and goal-driven generation steps, we design a prompt for GPT4 (fig.~\ref{fig:active_ingredient_extraction_prompt}).

\begin{figure*}[htbp]
\noindent\rule{\textwidth}{0.4pt}
\begin{lstlisting}
[System Message]
Reply with a succinct (i.e., 15 words or less) description of the following biological mechanism's active ingredient. Follow the instructions.
[Instructions]
- The active ingredient should describe how the species "act" upon its challenges to mitigate them, and include verb or verb phrasees.
- Active ingredient descriptions should also focus on the integral ingredients such as its bodily parts, liquids, or evolutionary tactic that are concrete and distinctive.
- Structure your output in the following format (do not output any characters other than the actual json-formatted dictionary):
{"desc": "..."}
\end{lstlisting}
\noindent\rule{\textwidth}{0.4pt}
\begin{lstlisting}
[User Message]
{mechanism}
\end{lstlisting}
\noindent\rule{\textwidth}{0.4pt}
\vspace{-1.5em}
\caption{The prompt used to extract the active ingredient from a mechanism. The mechanism is description to extract from is provided as part of the user message to GPT4.}
\label{fig:active_ingredient_extraction_prompt}
\end{figure*}

\subsection{Sparks}
\subsubsection{Spark Generation Prompt} \label{appendix:spark_generation}
The prompt used for generating sparks is detailed in fig.~\ref{fig:spark_generation_prompt}.
\begin{figure*}[htbp]
\noindent\rule{\textwidth}{0.4pt}
\begin{lstlisting}
[System Message]
Generate **2** highly different ideas that could solve the design problem: "{design_prob}".
The design problem has constraints that the ideas must satisfy: {design_constraints}
Generated ideas must be at least broadly related to the user-selected inspiration found in nature.
Generated ideas should be novel and not redundant with the following ideas generated in the past: {prev_sparks}
Describe each idea succinctly (i.e., max 500 characters), but ensure to provide sufficient details to help the user visualize the idea.
Start each idea description with a short, eye-catching name that captures the gist.
Output exactly in the following format, WITHOUT ANY OTHER TEXT:
[{{"name": "IDEA 1 NAME", "desc": "IDEA 1 DESCRIPTION"}}, ...]
\end{lstlisting}
\noindent\rule{\textwidth}{0.4pt}
\begin{lstlisting}
[User Message]
User-selected inspiration from nature to base your generation on:
===
{user_selected_mechanism_inspiration}
===
\end{lstlisting}
\noindent\rule{\textwidth}{0.4pt}
\vspace{-1.5em}
\caption{The prompt used to generate new sparks. We contextualize the prompt using the design problem description and the constraints provided with the problem, as well as 20 previously generated sparks for precedent-based diversification. We explicitly instruct the model to generate non-redundant sparks based on the history of precedents, and be succinct (\ie under 500 characters), with a descriptive title. Finally the user-selected mechanism inspiration is provided as part of the user message to GPT4.}
\label{fig:spark_generation_prompt}
\end{figure*}

\subsubsection{Precedent-based Diversification}
\label{appendix:precedent_based_diversification}
\begin{figure*}[ht]
    \begin{subfigure}[t]{.21\linewidth}
        \centering
        \includegraphics[height=4.5cm]{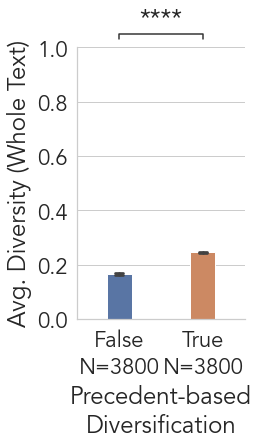}
    \end{subfigure}
    \begin{subfigure}[t]{.24\linewidth}
        \centering
        \includegraphics[height=4.5cm]{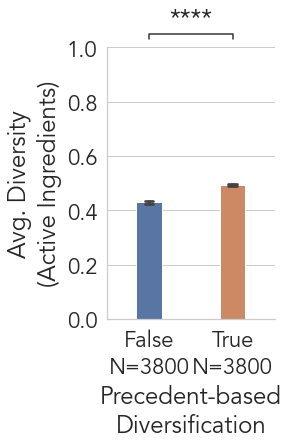}
    \end{subfigure}
    \quad
    \begin{subfigure}[t]{.24\linewidth}
        \centering
        \includegraphics[height=4.5cm]{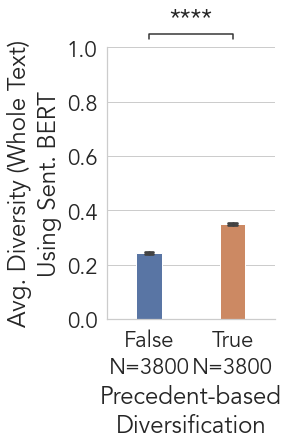}
    \end{subfigure}
    \begin{subfigure}[t]{.24\linewidth}
        \centering
        \includegraphics[height=4.5cm]{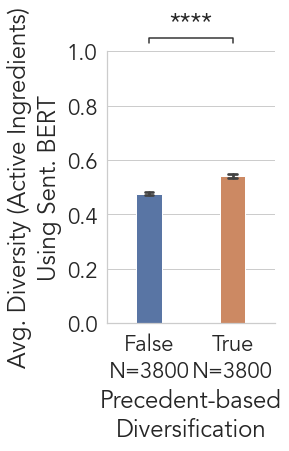}
    \end{subfigure}
    \vspace{-1em}
    \caption{(First \& Second) Bar graphs show that semantic diversity increased when using the precedent-based diversification approach, both at the whole spark and active ingredient levels; (Third \& Fourth) Repeat analyses show the robustness of these results against the choice difference of the encoder model, when the {Sentence-bert} model~\cite{reimers2019sentence} is used instead of OpenAI's {text-embedding-3-large}.}
    \label{fig:precedent_diversification}
    \vspace{-1em}
\end{figure*}
We tested whether this precedent-based diversification approach leads to a significant improvement in terms of semantic diversity compared to generation without diversification by generating 20 sparks for each of 10 randomly selected seed mechanism inspirations for each of the two design problems in the user study. 

We investigate semantic diversity at two levels, the whole text and the active ingredient of a spark.
To get the active ingredient, we process the generated spark using the same process as before for extracting active ingredients from mechanisms (\S\ref{subsubsection:new_backend_active_ingredient_extraction}).
We then encode each spark or active ingredient text into an embedding using the OpenAI's \codett{text-embedding-3-large} model.
We construct pairs of spark or active ingredient embeddings using the 20 sparks generated for each seed mechanism for each of the two design problems, which amounted to 3,800 pairs, and calculate the average cosine distance among the pairs.
This average represents the semantic diversity measure, which has been used in similar context in prior studies and was shown to be a viable measure of semantic diversity of natural language texts (\cf~\cite{gero2022sparks,hayati2023far,tevet2020evaluating}).
In order to ensure robustness of our results against the choice difference of the encoder model, we repeat the analysis using another popular encoder -- the \codett{Sentence-bert} model for embedding the text~\cite{reimers2019sentence}.
We find that, at the whole spark text level, the semantic diversity was significantly higher when precedent-based diversification was used (M=.24, SD=.073) than not (M=.17, SD=.090) (\tind{7291.87}{-42.41}{<< .0001}).

\subsection{Trade-off Analysis Generation} \label{appendix:trade_off_generation}
The prompt used for generating a trade-off analysis is detailed in fig.~\ref{fig:tradeoff_analysis_prompt}.
\begin{figure*}[htbp]
\noindent\rule{\textwidth}{0.4pt}
\begin{lstlisting}
[System Message]
Generate up to **3** anticipated pros and cons for applying the user-selected mechanism to the design problem: "{design_prob}".
The design problem has constraints that the ideas must satisfy: {design_constraints}
Format the 'pros' and 'cons' into each column in a markdown table.
Place the header row at the top of the table: "| **PROS** | **CONS** |".
After the header row, place each 'pro'-'con' row; start each 'pro' or 'con' text with a succinct label (3 words or less), enclosed in parantheses.
\end{lstlisting}
\noindent\rule{\textwidth}{0.4pt}
\begin{lstlisting}
[User Message]
User-selected inspiration from nature to base your generation on:
===
{user_selected_mechanism_inspiration}
===
\end{lstlisting}
\noindent\rule{\textwidth}{0.4pt}
\vspace{-1.5em}
\caption{The prompt used to generate a new potential design trade-off analysis. We contextualize the prompt using the design problem description and the constraints provided with the problem. We instruct the model to return the `pros' and `cons' of the mechanism inspiration in the context of the design problem using a markdown table format that places each pro-and-con pair in a new row, and give each item in the table a succinct (3 words or less) label. Finally the user-selected mechanism inspiration is provided as part of the user message to GPT4.}
\label{fig:tradeoff_analysis_prompt}
\end{figure*}

\subsection{Q\&A Response Generation} \label{appendix:q_and_a}
The prompt used for generating a response to the free-form user question is detailed in fig.~\ref{fig:q_and_a_prompt}.
\begin{figure*}[htbp]
\noindent\rule{\textwidth}{0.4pt}
\begin{lstlisting}
[System Message]
Reply to the following user question about the mechanism: "{inspos}".
Contextualize your response with the design problem: {design_prob}
And with the constraints of the problem: {design_constraints}
\end{lstlisting}
\noindent\rule{\textwidth}{0.4pt}
\begin{lstlisting}
[User Message]
User question about the mechanism:
===
{user_question}
===
\end{lstlisting}
\noindent\rule{\textwidth}{0.4pt}
\vspace{-1.5em}
\caption{\revised{}{The prompt used to generate a response to a free-form user question. We contextualize the prompt using the design problem description, its constraints, and the mechanism the user is asking about.}}
\label{fig:q_and_a_prompt}
\end{figure*}

\subsection{Detailed Research on Perplexity.ai}
\label{appendix:subsection_perplexity}
In order to support users with drilling down on related scientific research for each mechanism inspiration on demand, we designed a designated button (The `See more details on Perplexity.ai\footnote{\url{https://www.perplexity.ai/}}' button, fig.~\ref{fig:modal}, \cirnum{C}).
Through interface pilots, we anticipated that the most common user workflow for drilling down on related research to be taking place after the user decides on a particularly interesting cluster for further consideration.
When designing the button, we initially considered its placement on each of the cluster cards in the main interface, but decided to move it to the cluster modal view in order to prevent clutter and support effective exploration of diverse design space on the main interface.
In addition, to support the streamlined exploration -- decision -- further research workflow, we specifically placed the button at the end of the extended mechanism description featured in the modal fig.~\ref{fig:modal}, \cirnum{C}).

We implemented the button's functionality as opening a new browser tab that contains search results of relevant research on the Perplexity.ai website.
The search query was pre-populated using the following template:
\begin{lstlisting}[style=code]
Give me relevant details about "[active ingredient]" commonly found in [species]
\end{lstlisting}
This functionality design was a compromise following our technical investigation at the time of development that showed the difficulty of implementing Perplexity.ai's search page results inside a native \codett{React.js} application interface\footnote{Perplexity.ai prohibits user requests that attempt to render its search results natively.} and the lack of API\footnote{\url{https://docs.perplexity.ai/}} support for evidence generation\footnote{Last tested on March 17th, 2024.}.

\begin{figure}[h]
    \centering
    \includegraphics[height=6cm]{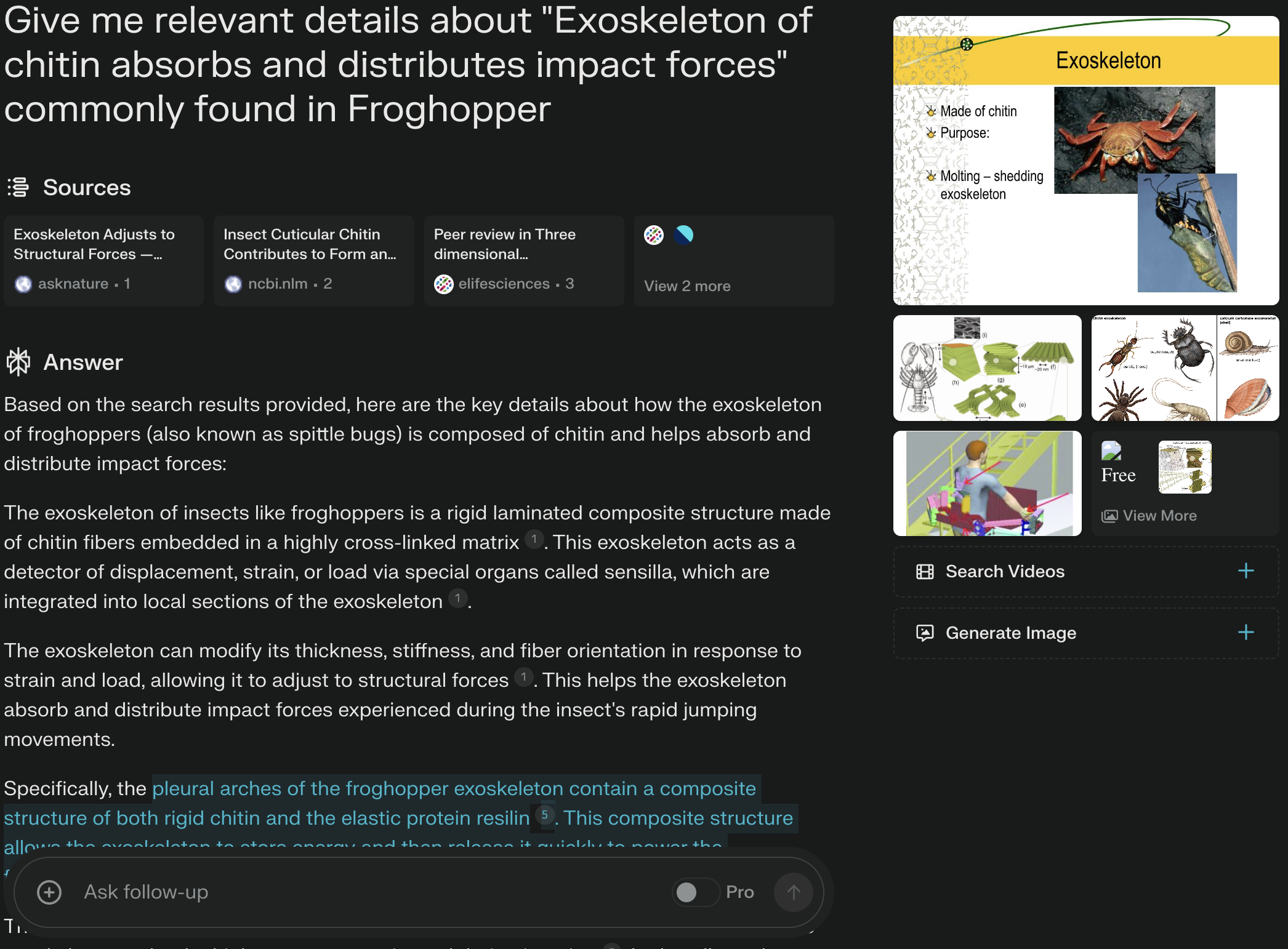}
    \vspace{-1em}
    \caption{An example Perplexity.ai result page in a new browser tab when the user clicks on the `See more details on Perplexity.ai' button on the mechanism modal view. The page describes how the froghopper exoskeleton contains a composite structure of both rigid chitin and the elastic protein resilin that allows the exoskeleton to store energy and then release it quickly to power the froghopper's powerful jumps, and its supporting research, that may provide valuable details as described in our scenario (\S\ref{subsubsection:walk_through}).}
    \vspace{-1em}
    \label{fig:pplx}
\end{figure}

\section{User Study Additional Details}
\subsection{Study System Tutorials}
\label{appendix:system_tutorials}
Before participants start with each of the two main task in each condition, they were given a tutorial of the assigned systems via screen sharing.
The interviewer demonstrated a step-by-step process and the main features of each system using a prepared script that took around 8 minutes for the \sys{} condition which had more features, and around 5 minutes for the baseline condition.
In the baseline condition, participants were instructed to open up 5 different URLs each pointing to a pre-curated list of mechanisms for a functional category.
The 5 functional categories used in the study were the same as those that were used for the \sys{} backend dataset pipeline, and they were: Manage Impact, Manage Tension, Manage Compression, Manage Turbulence, and Modify Speed.
In addition, participants in the baseline condition were instructed to sign in and open ChatGPT, and freely use it for understanding and ideating for the design problems using the information found from AskNature.
When participants came up with each new idea during the task, they were told to write it down in a prepared Google spreadsheet that was shared in the beginning of the task.
In the \sys{} condition, participants were told to keep the stream space as a holder for their ideas, and thus delete any ideas they did not like or edit the text directly.

\subsection{Examples of Judges' Novelty, Feasibility, Value Evaluation} \label{appendix:judge_scores_rationale_samples}
The following tables (Table~\ref{},\ref{}) present 8 examples of the judges' evaluations of design ideas in terms of novelty (N), feasibility (F), and value (V), along with the corresponding rationales. The evaluations were conducted independently by a domain expert and the first author using an agreed-upon rubric, demonstrating high reliability across these dimensions.
\begin{table*}[t]
\centering
\begin{tabular}{>{\small}p{4cm}|p{.2cm}|p{.2cm}|p{.2cm}|>{\small}p{3.3cm}|>{\small}p{4cm}|>{\small}p{3cm}}
\hline
\textbf{Formatted Ideas} & \textbf{N} & \textbf{F} & \textbf{V} & \textbf{Novelty Rationale} & \textbf{Feasibility Rationale} & \textbf{Value Rationale} \\
\hline
Idea 38: Inspired by the protective, retractable nature of turtle shells, this bike rack features a hard, aerodynamic shell that encases bikes completely. Its surface is segmented like scutes, allowing it to expand or contract to snugly fit 16", 20", and 26" bikes. When not in use or when driving, the shell retracts, minimizing drag and protecting bikes from theft and weather. & 8 & 6 & 8 & Connection to the turtle shell geometry is novel. The focus on the particular property of turtle shells which is that there is an empty space inside the shell where limbs can be retracted/folded into, and applying this design principle might provide useful new ideas for shape-shifting bike racks when the car is moving vs. stationary. & The estimated feasibility of the idea is somewhat high to start because the main mechanisms are extraction or contraction and segmented surface panels, which for the most part feel manufacturable with current tech and resources. However, more concrete design exploration, material selection, components design etc. may reveal previously unknown engineering challenges (hence the score is somewhat lowered for that reason). & The aerodynamic shape with expandable and contractable segmented scutes will be valuable for novel bikerack designs, and it opens up a new design space for future ideas. It may also be applicable to other domains where air or water resistance need to be considered, raising its value. \\
\hline
Idea 2: Tortoise Shell Geometry - For the bike racks stability \& durability, I would use layers of different properties to add on or improve the strength mimicing the tortoise shell geometry. Also I would adapt this feature since it provides very good longevity of the product which enhances user satisfaction \& trust. & 6 & 3 & 4 & In contrast to the earlier turtle shell idea that focused on the specific retraction mechanism, this one focuses more on the shape and geometry of the shell. However, the ``use layers of different properties'' is not very specific or novel. & ``Use layers of different properties to add on or improve the strength mimicing the tortoise shell geometry'' sounds like it would require a lot of materials research and feasibility testing before making this idea feasible and useful. & For durability, an efficient shell geometry that can be mass produced could be valuable. However it is unclear how multiple bike sizes will be accommodated. Hence the value is somewhat low. \\
\hline
Idea 18: Rack should be top mounted so that the bike is facing the same direction as the car to minimize drag (i.e. least amount of surface exposed to air during the ride)	& 1	& 10 & 2 & ``Top mounted rack'' already exists and does not offer much novelty in the idea.	& Rack mounted atop and facing the same direction as the car already exists as a commercial product, hence high engineering feasibility of the idea.	& Top mounted racks are valuable, but the idea already being commercially available, unclear how much additional value (of any) this idea adds. Furthermore, it does not specify how the air resistance could be further reduced nor how the solution will accommodate different bike sizes, hence the value of the idea is low.\\
\hline
Idea 79: LeapLock FlexFit. Inspired by Anura's powerful legs, this bike rack uses a bio-mimetic spring mechanism that stretches to accommodate 16", 20", and 26" bikes. Its skeletal structure, mimicking frog bones, flexes to absorb road vibrations, protecting bikes. Aerodynamically shaped to reduce drag, it 'leaps' into a compact form when not in use, preserving fuel efficiency.	& 9	& 6	& 9	& The spring mechanism that changes shape when in use vs. not, and also being able to absorb vibration and impact feels like a pretty novel idea. & Springs already exist, and there is presumably a large body of research around them that this idea could draw upon. However, the specific kinds of springs that exhibit the right kinds of properties here (like stretching the right amounts to accommodate for different bike sizes) might require a non-trivial amount of additional research into the materials and the construction of the springs etc. & This kind of novel spring mechanism that is aerodynamically shaped and size-adjustable could be highly valuable.\\
\hline
\end{tabular}
\caption{Sample judges' evaluations of design ideas by novelty (N), feasibility (F), and value (V), with rationales (continued in Table~\ref{tab:sample_judges_eval_2}).}
\vspace{-3em}
\label{tab:sample_judges_eval_1}
\end{table*}

\begin{table*}[t]
\centering
\begin{tabular}{>{\small}p{3.1cm}|p{.2cm}|p{.2cm}|p{.2cm}|>{\small}p{3.8cm}|>{\small}p{3.8cm}|>{\small}p{3.8cm}}
\hline
\textbf{Formatted Ideas} & \textbf{N} & \textbf{F} & \textbf{V} & \textbf{Novelty Rationale} & \textbf{Feasibility Rationale} & \textbf{Value Rationale} \\
\hline
Idea 111: Drawing from the compact and sturdy design of a beetle's shell, this bike rack folds out from a sleek, aerodynamic shell attached to the car's roof. It adjusts to fit various bike sizes, using a secure, adjustable locking mechanism that mimics the versatility of beetle wings. & 8 & 5 & 7 & The connection to the beetle's shell is novel. The connection between the beetle wings and adjustable locking mechanisms is also quite novel. It feels there is a fairly large gap of research from beetle wings to adjustable locking mechanisms, hence unclear whether current tech can manufacture it. adjustable locking mechanism could be valuable for accommodating different bike sizes and dynamically changing volume/shape for aerodynamic efficiency. The expandable shell attached to the roof of a vehicle is also a somewhat valuable idea. & It feels there is a fairly large gap of research from beetle wings to adjustable locking mechanisms, hence unclear whether current tech can manufacture it. & adjustable locking mechanism could be valuable for accommodating different bike sizes and dynamically changing volume/shape for aerodynamic efficiency. The expandable shell attached to the roof of a vehicle is also a somewhat valuable idea. \\
\hline
Idea 18: Dragonfly/Damselfly wing structure consider structure for titanium frame & 5 & 8 & 3 & The focus on the dragonfly wing structure seems novel. But the idea lacks substantive details as to what exactly about the wing structure is it that would make the frame structure better. & There is a decent amount of omitted details for which additional research may be required to fill the gap. I think the overall wing pattern may be something that current manufacturing could replicate in a relatively straightforward manner. & This idea alone does not directly address the challenge of wheelchairs going up the stairs. It is unclear how applying the dragonfly wing structure to the frame (hence perhaps making the frame sturdier?) will help wheelchair designs to move up the stairs. The wing structure-inspired frame may be light and durable, which is desirable and meets one of the specified problem constraints, and could be used in combination with other mechanisms that actually solve the main challenge of going up the stairs?\\ \hline
Idea 66: Inspired by tortoises, this wheelchair features a retractable design where the seat, wheels, and footrest fold into a durable, hard shell case. This shell is made from lightweight, impact-resistant materials, ensuring protection during transit. For stair climbing, it utilizes a set of extendable tracks that grip onto steps, smoothly elevating the user.  & 7 & 7 & 7 & The focus on the retractability of the shell is novel and distinguishable from a more generic variation of the idea that would say something like "apply the shell-like geometry". For climbing, it mentions the use of extendable tracks that grip on to steps. This part perhaps isn't as novel in light of existing wheelchair ramps and other track-based mechansisms. & Each individual parts of the solution seems fairly feasible given the existing similar mechanisms in other commercial products (shell-like geometry in backpacks or clothing, like extendable tracks in some other products...) However, producing an elegant design solution that combines the different aspects of the idea might be difficult and less feasible. & The retractable design with extendable tracks design would be valuable assuming they work well with different types of stairs while also foldable and durable.\\ \hline
Idea 15: German engineering- iStruct robot as a model for wheelchair & 8 & 5 & 6 & This idea reimagines wheelchair design using a robot, which feels like a provocative idea. & The feasibility assessment would require much more specification of how the iStruct robot would actually work as a wheelchair  and various other individual sub-problems related to that to make them into a coherent whole. It feels like there is a substantial amount of uncertainty about what exactly the idea is, and how the iStruct robot would be adapted into a feasible engineering solution. & Robotic design would be useful beyond just stair climbing and has potential to address challenges of any rough terrain types, however it would not be as foldable or lightweight or compact. \\ \hline
\end{tabular}
\caption{Additional sample judges' evaluations of design ideas by novelty (N), feasibility (F), and value (V), with rationales.}
\label{tab:sample_judges_eval_2}
\end{table*}

\subsection{Study Result: Top-5 and Bottom-5 Ideas, Scores, and Coding Analogical Transfer}
\label{appendix:top5_bottom5_ideas}
See Table~\ref{tab:representative_ideas}.
\begin{table*}[t]
\centering
\begin{tabular}{clcccp{11cm}}
    \hline
    \textbf{Rank} & \textbf{Cond.} & \textbf{Prob.} & \textbf{Score} & \textbf{An-Tr?} & \textbf{Idea} (Summarized) \\
    \hline
    T1 & \sys{} & 2 & 7.86 & Y &  {\small This bike rack uses a \textbf{bio-mimetic spring mechanism}, inspired from \textbf{Anura's powerful legs}, to stretch (and compress) to accommodate different bike sizes. Its skeletal structure, mimicking frog bones, flexes to absorb road vibrations, protecting bikes. Aerodynamically shaped to reduce drag, it `leaps' into a compact form when not in use, preserving fuel efficiency.} \\
    T2 & \sys{} & 2 & 7.62 & Y & {\small \textbf{Ultrasonic sensors to adjust its grip} on different bike frames automatically, maybe something like \textbf{cetaceans' echolocation}. As a bike is loaded, sensors emit sound waves that measure the frame's size, adjusting the rack's arms for a perfect fit without manual intervention. Its sleek, aerodynamic design mimics a beluga's streamlined shape, minimizing air resistance.}\\
    T3 & \sys{} & 2 & 7.62 & Y & {\small The bike rack utilizes \textbf{smart materials to morph its surface} to perfectly fit different bike sizes, inspired by the \textbf{changing shapes of leaves}. It's mounted on the sedan's roof, its leaf-like design folds to minimize air resistance, enhancing aerodynamics. When bikes are mounted, it expands and shapes itself to the contours of each bike, ensuring a secure fit for 16", 20", and 26" frames without adapters, then folds back seamlessly when not in use.}\\
    T4 & \sys{} & 2 & 7.49 & Y & {\small Bitterns' balance in currents $\rightarrow$ A dynamic, flexible support system.}\\
    T5 & \sys{} & 2 & 7.47 & Y & {\small Tree frogs' adhesive toe pads $\rightarrow$ a sticky, yet non-residual, surface-based attachment.}\\
    \hline
    B5 & \sys{} & 1 & 5.07 & Y & {\small Serpentine movement $\rightarrow$ Wheelchair with a flexible, segmented moving base.}\\
    B4 & \sys{} & 2 & 5.01 & Y & {\small Diatoms' protective frustules $\rightarrow$ Rear-mounted rack w/ adjustable, silica-patterned arms.}\\
    B3 & \sys{} & 1 & 4.93 & Y & {\small Froghopper's shock-absorbent exoskeleton $\rightarrow$ Lightweight \& foldable material.}\\
    B2 & \sys{} & 2 & 4.48 & N & {\small Bikes are `locked' in place by a vacuum seal mechanism.}\\
    B1 & \sys{} & 1 & 3.98 & Y & {\small Retractable `legs'; Jointed segments that extend and contract to climb.}\\
    \hline
    T1 & Baseline & 2 & 6.80 & Y & {\small \textbf{Aerodynamically shaped bikerack} design inspired by how \textbf{marine mammals like dolphins and whales} have evolved flippers and tail fins that optimize their movement in water. Their body shape tends to be more rounded than that of fish but is streamlined for efficient travel. The tail fins (flukes) provide powerful propulsion, while the pectoral flippers are used for steering and stabilization.}\\
    T2 & Baseline & 2 & 6.49 & Y & {\small A series of \textbf{adjustable, branching arms} extend gracefully from the main body, inspired by the natural \textbf{branching patterns of trees}. These arms adjust intuitively to hold various bike frames snugly, replicating the way branches support varying weights and shapes.}\\
    T3 & Baseline & 2 & 6.49 & Y & {\small \textbf{The rack's ``skin'' mimics shark denticles}, promoting laminar flow and drastically reducing drag. It's not only about fuel efficiency; it's about harmonizing with the very essence of movement.}\\
    T4 & Baseline & 1 & 6.33 & N & {\small Compressing and expanding mechanism for stairs; multiple wheel contacts.}\\
    T5 & Baseline & 2 & 6.32 & Y & {\small Marine polychaete worm-inspired design with tentacle-like grips for bikes.}\\
    \hline
    B5 & Baseline & 1 & 2.82 & N & {\small Wheelchair self-repair using Memory Shape Alloy.}\\
    B4 & Baseline & 1 & 2.82 & N & {\small Multiple contact points for wheelchair wheels.}\\
    B3 & Baseline & 1 & 2.71 & N & {\small Crystal sensors for damage detection and vital monitoring.}\\
    B2 & Baseline & 2 & 2.71 & N & {\small Top-mounted rack, bike aligned with car for aerodynamics.}\\
    B1 & Baseline & 1 & 2.62 & N & {\small Golden bamboo material for durability and lightweight design.}\\
    \hline
    \end{tabular}
    \caption{Top-5 and Bottom-5 scoring ideas from each condition, and they exhibit analogical transfer. `Rank': T1 -- T5 correspond to Top-1 -- Top-5 and B5 -- B1 to Bottom-5 -- Bottom-1 rankings; `Cond.' represents the condition in which the idea was produced; `Prob.' represents the problem the idea is for (1: the `wheelchair' problem; 2: the `bike rack' problem); `Score' represents the geometric mean of expert-judged novelty, value, and feasibility scores; `An-Tr?' is a binary value representing the presence of Analogical Transfer; `Idea' is summarized for conciseness, except for the top-3 ideas that were presented in verbatim. Boldfaced text in the verbatim ideas represents the source and the target in analogical transfer.}
    \label{tab:representative_ideas}
\end{table*}

\subsection{Study Data Analysis: Extraction of unique design constraints described in each idea} \label{appendix:study_extract_constraints}
To analyze the user engagement patterns involving consideration of design constraints (\S\ref{subsubsection:addressing_design_constraints}), we first extract the unique design constraints described by participants in each idea.
We use GPT4 (\codett{gpt-4-turbo-preview}) with a prompt (fig.~\ref{fig:constraints_extraction_prompt}) to perform the extraction.
The first author reviewed the extracted constraints in terms of their coherence and uniqueness for a random set of \codett{20} ideas and found that the extraction was satisfactory in terms of both the uniqueness of extracted constraints and their coherence.
Using this data, we find that the length of each idea was also significantly correlated with the number of design constraints described in it (\pearson{.58}{ < .0001}) (see \S\ref{subsubsection:addressing_design_constraints} for the related analysis).
\begin{figure*}[htbp]
\noindent\rule{\textwidth}{0.4pt}
\begin{lstlisting}
[System Message]
We're preparing each idea description for measuring the degree to which various design constraints were considered.

Chunk the idea description into unique segments each of which describe consideration(s) for a single coherent design constraint.
Output exactly in the following format (use double quotation marks to encapsulate any string), without any other text:
{"constraint_considerations": [["idea 1 constraint 1", "idea 1 constraint 2", ...], ... }
\end{lstlisting}
\noindent\rule{\textwidth}{0.4pt}
\begin{lstlisting}
[User Message]
{list_of_participant_ideas}
\end{lstlisting}
\noindent\rule{\textwidth}{0.4pt}
\vspace{-1.5em}
\caption{The prompt used to extract coherent chunks of text that relates to unique design constraints. Participants' ideas are stringified and provided as part of the user message to GPT4.}
\label{fig:constraints_extraction_prompt}
\end{figure*}

\subsection{Study Data Analysis: Extraction of the species' names that participants described as inspiring their ideas} \label{appendix:species_extracton}
To analyze the diversity of the species that participants were inspired from for their own ideas, we use GPT4 (\codett{gpt-4-turbo-preview}) with a prompt (fig.~\ref{fig:species_extraction_prompt}) to extract the species name from each participant idea and normalize it.
The first author then reviewed the extracted species' names in a random sample of \codett{20} ideas and found that the extraction accuracy was satisfactory.
\begin{figure*}[htbp]
\noindent\rule{\textwidth}{0.4pt}
\begin{lstlisting}
[System Message]
We're extracting the source species' name that inspired each idea from the idea description.

Output exactly in the following format (use double quotation marks to encapsulate any string), without any other text:
{"source_species": ['species name for idea 1', 'species name for idea 2', ...] }
\end{lstlisting}
\noindent\rule{\textwidth}{0.4pt}
\begin{lstlisting}
[User Message]
{list_of_participant_ideas}
\end{lstlisting}
\noindent\rule{\textwidth}{0.4pt}
\vspace{-1.5em}
\caption{The prompt used to extract the species' names that inspired participants' ideas. Participants' ideas are stringified and provided as part of the user message to GPT4.}
\label{fig:species_extraction_prompt}
\end{figure*}

\end{document}